\documentclass{emulateapj}

\usepackage{amsmath}
\usepackage{amstext}
\usepackage{units}
\usepackage[caption=false]{subfig}

\renewcommand{\ion}[2]{#1\,{\sc #2}}

\shorttitle{Transition from Magellanic Stream to Halo}
\shortauthors{Nigra et al.}
\submitted{Accepted for publication in The Astrophysical Journal}

\begin{document}

\title{Probing the Structure and Kinematics of the transition Layer between
the Magellanic Stream and the Halo in H\,\textsc{i}}

\author{Lou Nigra\altaffilmark{1}, Sne\v{z}ana Stanimirovi\'{c}, John
S. Gallagher, III }

\affil{Department of Astronomy, University of Wisconsin, 475 North Charter
Street, Madison, WI 53706, USA; }

\email{Electronic: lou@zooniverse.org}

\author{Kenneth Wood\altaffilmark{2}}

\affil{School of Physics and Astronomy, University of St. Andrews, North
Haugh, St. Andrews KY16 9SS, UK; }

\and{}

\author{David Nidever, Steven Majewski}

\affil{Department of Astronomy, University of Virginia, Charlottesville,
VA, 22904-4325, USA; }

\altaffiltext{1}{Citizen Science Department, Adler Planetarium, 1300 South Lake Shore Drive, Chicago IL 60605, USA}
\altaffiltext{2}{Department of Astronomy, University of Wisconsin Madison, 475 North Charter Street, Madison, WI 53706, USA} 
\begin{abstract}
The Magellanic Stream (MS) is a nearby laboratory for studying the
fate of cool gas streams injected into a gaseous galactic halo. We
investigate properties of the boundary layer between the cool MS gas
and the hot Milky Way halo with 21 cm \ion{H}{i} observations of
a relatively isolated cloud having circular projection in the northern
MS. Through averaging and modeling techniques, our observations obtained
with the Robert C. Byrd Green Bank Telescope (GBT), reach unprecedented $3\sigma$
sensitivity of $\sim1\times10^{17}$ cm$^{-2}$, while retaining the
telescope's $9.1\,'$ resolution in the essential radial dimension.
We find an envelope of diffuse neutral gas with $\text{FWHM}\text{ of }60\,\text{km}\,\text{s}^{-1}$,
associated in velocity with the cloud core having $\text{FWHM}\text{ of }20\,\text{km}\,\text{s}^{-1}$,
extending to 3.5 times the core radius with a neutral mass seven times
that of the core. We show that the envelope is too extended to represent
a conduction-dominated layer between the core and the halo. Its observed
properties are better explained by a turbulent mixing layer driven
by hydrodynamic instabilities. The fortuitous alignment of the \objectname{NGC 7469}
background source near the cloud center allows us to combine UV absorption
and \ion{H}{i} emission data to determine a core temperature of
$8350\pm350\,\text{K}$. We show that the \ion{H}{i} column density
and size of the core can be reproduced when a slightly larger cloud
is exposed to Galactic and extragalactic background ionizing radiation.
Cooling in the large diffuse turbulent mixing layer envelope extends
the cloud lifetime by at least a factor of two relative to a simple
hydrodynamic ablation case, suggesting that the cloud is likely to
reach the Milky Way disk.
\end{abstract}


\section{\label{sec:Introduction}Introduction}

The Magellanic Stream (MS) is our closest and most prominent example
of a gaseous interaction remnant. While such circumgalactic structures
are postulated to represent important sources of fuel for future star
formation \citep{keres05,wakker08,dekel05,brooks09}, the mechanisms
whereby this material might be accreted back into galaxies remain
unclear. As emphasized by \citet{keres05}, the multiphase nature
of galactic halos plays an important role during the accretion process
by modifying gas stripping and infall processes \citep[for additional perspectives on these issues see ][]{silk87,gallagher05,tullman06}.
Being close by, the MS offers a unique laboratory to study the rate
and nature of gas injected from satellites, as well as models to assess
the fate of the stripped gas.

The MS trails across much of the southern Galactic sky behind the
Magellanic Clouds (MCs), passing near the Southern Galactic Pole at
about 1/3 of its length. It has been shown to have a continuous coherent
velocity tracing across the southern Galactic sky for $140^{\circ}$
\citep{nidever10}, just penetrating the Galactic plane at some unknown
distance from the Galactic center. Direct distance constraints for
the MS are difficult to establish much beyond the MCs themselves,
which constrain the MS head nominally between $52$ and $61$~kpc,
based on the MC distances established by \citet{koerwer09} and \citet{hilditch05}.
With no embedded stars, no nearby ionizing sources or stellar absorption
sight lines yet detected, the MCs remain its only direct distance
constraint. A geometrical-dynamical analysis by \citet{jin08} found
a Galactocentric distance that increased to $70$~kpc at about $90^{\circ}$
along the MS length. \citet{stani08} found a similar constraint at
this point assuming that the neutral hydrogen (\ion{H}{i}) clump
size distribution is driven by the Thermal Instability (TI).

Attempts to simulate the large-scale mechanism of formation and subsequent
evolution of this extensive structure have focused mainly on tidal
interactions with the Galaxy and/or ram pressure stripping of the
MC gas by the Galactic Halo. These simulations had moderate success
when it was believed that the MCs have had multiple orbital cycles
around the Galaxy \citep{moore94,mastro05,connors06}. However, global
simulations have been made difficult in recent years given the relatively
new constraint that the MCs are apparently on their first pass through
the Galactic system \citep{besla07}. \citet{besla10} developed a
Smoothed Particle Hydrodynamics (SPH) global MS simulation in this
first-pass scenario maintaining that neither tidal or ram pressure
mechanisms are as sufficient as once thought to stripping the gas.
Both \citet{besla10} and \citet{nidever08} propose other mechanisms
to do the heavy lifting, unbinding the gas prior to stripping, through
tidal resonances between the MCs in the former, and in the latter,
a gas ``blowout'' from star formation in the LMC region that includes
30~Doradus.

\begin{figure*}
\begin{centering}
\includegraphics{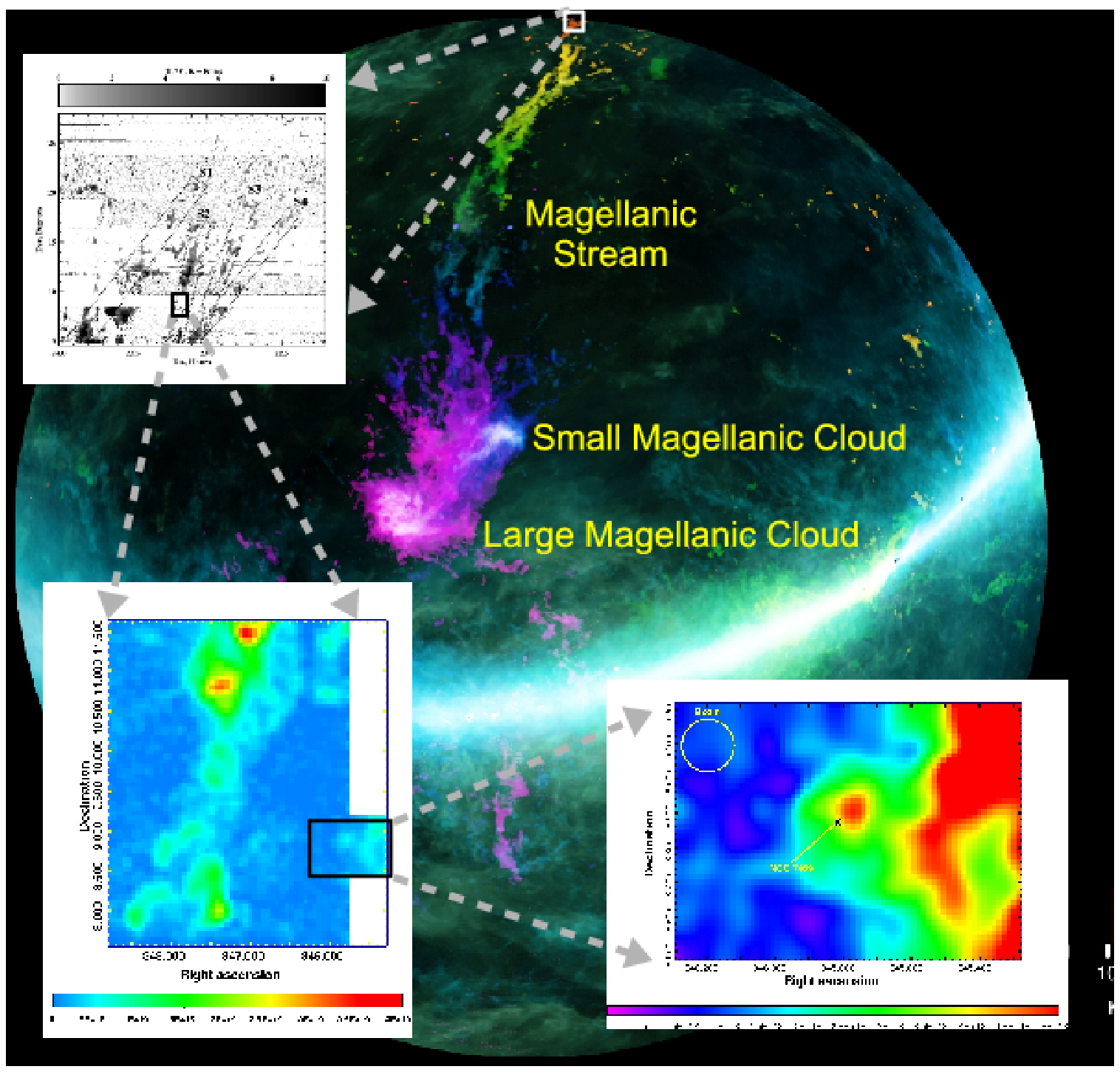} 
\par\end{centering}

\caption[A journey in scales.]{A journey in scales: From the entire MS, illustrated in the southern
Galactic hemisphere map from the Galactic All Sky Survey \citep[GASS, ][]{mcclure09}),
down to the specific cloud studied in this paper. Counterclockwise
from top: High resolution GALFA Arecibo Observatory \ion{H}{i}
map revealing fine filamentary structure in the northern MS \citep{stani08},
high sensitivity GBT map of a section of one of those filaments (this
work), and the cloud discovered in that map containing the NGC 7469
sight-line.\label{fig:MStoClump}}
\end{figure*}

Whatever the mechanism by which the gas was removed from the MCs,
ram pressure, tidal effects, initial turbulent energy, magnetic field
structure and subsequent interactions with the Galactic Halo all may
have played a role in sculpting the rich filamentary and clumpy structure
revealed in Figure~\ref{fig:MStoClump}. 
This figure shows the scale of the entire MS on an \ion{H}{i} image
produced by the Galactic All-Sky Survey \citep[GASS,][]{mcclure09}
and takes it down in steps to a specific small, but interesting cloud
toward the downstream end which is the subject of this paper. The
fine scale structure revealed in the GASS image and in other recent
\ion{H}{i} observations such as \citet{putman03a} and \citet{bruns05}
is not captured in the N-Body and SPH global models mentioned above.
Even finer structure was revealed by \citet{stani02,stani08} in high
resolution mapping of the northern tip of the MS using the 305~meter
dish at Arecibo. These observations revealed extended fine filamentary
structure and clumps down to the $3.5'$ angular resolution of the
telescope.

Several theoretical studies as well as numerical simulations have
explored the effect of various hydrodynamic instabilities on the formation
of small scale structure in warm gas such as the MS moving through
a hot halo \citep{mori01,quilis01,heitsch09}. However, it is still
not clear how effective these processes are and on what timescales
they operate. For example, most instabilities appear to act on short
timescales ($\sim100$~Myrs) and therefore require that the MS is
continuously replenished with fresh material. The nature and evolution
of the small scale MS structure (or stripped gas) is also not well
understood. As such material may eventually constitute a substantial
source for the MW's star formation in the form of infalling warm ionized
gas \citep[so called ``warm drizzle'',][]{bh07}, observational constraints
on the physical processes operating on small spatial scales in the
MS are highly important.

Numerous observational studies have revealed a multi-phase nature
in MS gas. \citet{kalberla06} and \citet{stani08} showed that about
15\% of the observed \ion{H}{i} clouds have velocity profiles composed
of warm and cool components, with a velocity FWHM of about 25 km~s$^{-1}$
and 3-15 km~s$^{-1}$, respectively. Matthews et al. (2009) detected
the first \ion{H}{i} absorption lines against radio background
sources in the direction of the MS close to the MCs indicating spin
temperature of 70-80 K. H$\alpha$ measurements of the warm ionized
component in the MS exist \citep{weiner96,putman03b}, but with only
a few discrete pointings and with the large $\sim1^{\circ}$ apertures,
providing little insight into the processes on the small, arc minute
scales.

While not being able to provide any spatial information, numerous
UV and optical absorption studies have been crucial for constraining
the abundance of the MS gas \citep{slavin93,fox10}. In addition,
detections of \ion{O}{vi} absorption from gas associated with the
MS by \citet{sembach03} give strong support for the existence of
an ionized component around the MS with $T<10^{6}$ K. It is generally
interpreted that this component represents an interface between the
hot Halo gas at $T\sim10^{6}$ K \citep{fang06} and the cooler MS
gas. Studies of lower ionization states suggest the existence of diffuse
envelopes of somewhat cooler, partially ionized gas that is not visible
in the current \ion{H}{i} surveys. Specifically, \ion{Si}{iii}
likely probes different phases than \ion{O}{vi} with $T=10^{4-4.5}$
K \citep{shull09} and has been detected along a sight-line in the
northern MS with velocities associated with the MS \citep{collins09}.
Recently, \citet{fox10} have performed an analysis of many low and
high ion species in UV and optical absorption, including \ion{Si}{iii},
against background source \objectname{NGC 7469}. From this work,
a picture emerges of a diffuse, multi-phase transition structure between
the warm, mostly neutral envelope gas detected in \ion{H}{i} and
the hot, mostly ionized envelope gas detected in \ion{O}{vi}.

As pointed out in \citet{stani10}, the rich multi-phase structure
of the MS suggests cloud longevity and a slow mass ablation rate.
The Kelvin-Helmholtz instability (KH) driven by shear flow between
cool MS gas and the hot Halo is likely to be the dominant mode of
cool cloud disruption and relatively rapid ablation compared to evaporation
through pure thermal conduction. However, analytical and numerical
treatments of cool gas in a hot flow suggest that the rapid ablation
by KH can be moderated by factors affecting the turbulent mixture
in the boundary layer. These include magnetic fields \citep{esquivel06},
re-cooling of heated gas \citep{begelman90,kwak11} and even thermal
conduction acting locally within the mixture \citep{vieser07a}.

The focus of this paper is to probe and characterize properties of
this boundary layer between cool \ion{H}{i} clouds in the MS and
the surrounding Halo gas. This requires very deep \ion{H}{i} observations,
more sensitive than the existing \ion{H}{i} surveys. For example,
the predominantly ionized component detected by \citet{collins09}
has the total column density of $\sim10^{18-19}\,\text{cm}^{-2}$
and a neutral fraction of $\sim0.01$. To detect the corresponding
neutral gas we require \ion{H}{i} column density sensitivity of
$\sim10^{16-17}\,\text{cm}^{-2}$. This is at least 5 times lower
than the most sensitive \ion{H}{i} survey to date; GASS \citep{mcclure09}
which achieves $3\sigma$ column density sensitivity of $1.6\times10^{18}\,\text{cm}^{-2}$.
We approach the required sensitivity by applying a new method of spatial
averaging to characterize a cloud in the northern MS (shown in Figure~\ref{fig:MStoClump})
observed with very deep \ion{H}{i} emission spectra obtained with
the Green Bank Telescope%
\footnote{The Robert C. Byrd Green Bank Telescope is operated by the National
Radio Astronomy Observatory, which is a facility of the US National
Science Foundation operated under cooperative agreement by Associated
Universities, Inc. %
} (GBT). Our analysis also benefits from the fortuitous location of
the \objectname{NGC 7469} Seyfert~1 background source near the cloud's
center which was extensively studied in UV and optical absorption
by \citet{fox10} (hereafter referred to as FSW10).

Our observing and data reduction strategies are presented in Section~\ref{sec:Observations}.
Section~\ref{sec:Results} explains the new spatial averaging method
we applied to the cloud to achieve $<10^{17}\,\text{cm}^{-2}$ sensitivity
and probe the usually inaccessible cloud periphery and highly ionized
cloud envelope. We also develop a simple 3D cloud model matched to
our observations to obtain a robust estimate of cloud's properties
and gain insight into the internal cloud structure and processes acting
at its periphery. In Sections~\ref{sub:Discussion:Envelope} and
\ref{sub:Discussion:Core}, we show that the cloud properties along
with results from FSW10 strongly suggest an integral WNM/WIM core
transitioning to a Turbulent Mixing Layer (TML) and in Section~\ref{sub:Discussion:Ionization},
we investigate how consistent the cloud's neutral density characteristics
are with a photo-ionizing environment by comparing the model to the
results of 3D ionization equilibrium simulations. Finally, in Section~\ref{sub:Discussion:Lifetime}
we discuss the lifetime of the cloud and assess the possibility of
eventually reaching the Galactic Disk where a portion of its mass
could help fuel star formation. We summarize our results in Section~\ref{sec:Conclusions}.

\section{Observations and Data reduction}

\label{sec:Observations}

We used the GBT in 2009 and 2010 as part of a larger program to map
several sample regions in the northern portion of the MS. One of these
regions is located in the northern tip of the MS and was chosen, among
other reasons, to include the NGC~7469 background source. The beam
size of the GBT in the 21 cm line is 9.1$'$, which corresponds to
183 pc at $70\,\text{kpc}$, a distance consistent with estimates
by \citet{jin08} and \citet{stani08} which we will assume hereafter.

Observations were obtained using the on-the-fly (OTF) mapping mode
with in-band frequency switching. Constant Declination (DEC) rows
were scanned and stepped every $3.5'$ in order to oversample the
beam by 2.6 in the DEC direction. Spectra were dumped at twice that
resolution so that beam-spreading in the Right Ascension (RA) direction
due to scanning remained small ($\sim2\%$). The $36$~second target
integration time per $3.5'$ pixel was split into four separate scans,
which reduced the impact of transient signals such as Radio Frequency
Interference (RFI) and spectrometer glitches. The GBT spectrometer
was used with a bandwidth of 12.5 MHz and 16384 channels, corresponding
to a velocity resolution of $0.161\,\text{km}\,\text{s}^{-1}$.

The GBT data were reduced using a combination of the GBTIDL and AIPS
data reduction packages, and a suite of specially designed Interactive
Data Language$^{\text{\textregistered}}$ (IDL) programs. To ensure
high sensitivity, special care was taken during the baseline calibration
process. First, a modified version of the GBTIDL GETFS procedure was
used to obtain reference spectra to produce temperature-calibrated
spectra with gross baseline calibration. Instead of using a frequency-shifted
total power reference spectrum or a smoothed version of it, as is
normally done, a noiseless 3rd order baseline model of a line-suppressed
total power reference spectrum was used for reference. This improves
the Signal-to-Noise Ratio (SNR) by $\sqrt{2}$ without enhancing large-scale
baseline structure noise which degrades its Gaussian character. The
line-suppressed spectrum was obtained by selecting the minimum of
either the reference or signal spectrum at each frequency. This ``min
filter'' statistically prefers samples without a positive bias from
emission of any kind and so reduces image artifacts in the final spectra
and adds a fixed offset to the baseline, easily removed in the following
steps. The derived reference spectrum was then used to perform a $(Signal-Reference)/Reference$
calibration of each spectrum. The spectra were then scaled by the
system temperature and corrected for atmospheric opacity using the
standard, built-in correction provided by GBTIDL routines.

In the next step, the calibrated and roughly baselined spectra pass
through two additional baseline removal phases. Each of these include
multiple steps where baselines are modeled and subtracted. In all
but one case, baselines are modeled as 3rd order polynomial functions
only after excluding Galactic emission and subtracting unblended Gaussian
models of any detected emission lines in the spectrum. In the first
phase, the baseline of the scan average for the X polarization is
modeled and then subtracted from each spectrum in that scan. The baselined
X emission is then subtracted from the Y polarization, effectively
suppressing all emission under the reasonable assumption in this case
that the emission is unpolarized. This leaves only the Y baseline,
which is modeled to 7th order. This special step removes the small
$\approx1.5$~MHz sinusoidal baseline component present in the Y
polarization of GBT spectra, probably due to double transit reflections
from the feed to the circumferential gap between main reflector panels
\citep{fisher02}. It also removes the average offset introduced by
the min filter mentioned previously. In the second step, each individual
spectrum's residual baseline differences from the average are removed
in two iterations.

After baselining, spectra from identical pointings were combined and
interpolated to a regular grid with pixel size one fourth that of
the sampling interval (or $0.875'$) using the AIPS task SDGRD. This
over-sampled the telescope beam by a factor of ten in the final data
cube, providing interpolated resolution to support precise averaging
along circular paths when spatial averaging techniques are applied
to the cube. Velocity spectra were smoothed with a 6th order Hanning
window, reducing the resolution to $0.966\,\text{km}\,\text{s}^{-1}$.
The velocity range was truncated to $-500$ to $-200\,\text{km}\,\text{s}^{-1}$.
The rms noise in the final data cube was measured at $\sigma_{T}=3.4$~mK
by averaging across emission-free velocity ranges. This corresponds
to column density noise $\sigma_{N}=1.0\times10^{17}\,\text{cm}^{-2}$
for a $15\,\text{km}\,\text{s}^{-1}$ FWHM profile. This is almost
four times better than the $3.8\times10^{17}\text{cm}^{-2}$ noise
achieved by GASS, the most sensitive survey to date \citep{mcclure09}
when scaled to the same FWHM line width of $15\,\text{km}\,\text{s}^{-1}$.
This sensitivity is provided at the GBT beamwidth of $9.1'$ compared
to the Parkes telescope's $14'$.

\section{Results: A Simple Core + Envelope Cloud Model}

\label{sec:Results}

\begin{figure*}
\begin{centering}
\subfloat[Column Density (color scale in $\text{cm}^{-2}$)]{\begin{centering}
\includegraphics[width=0.35\paperwidth]{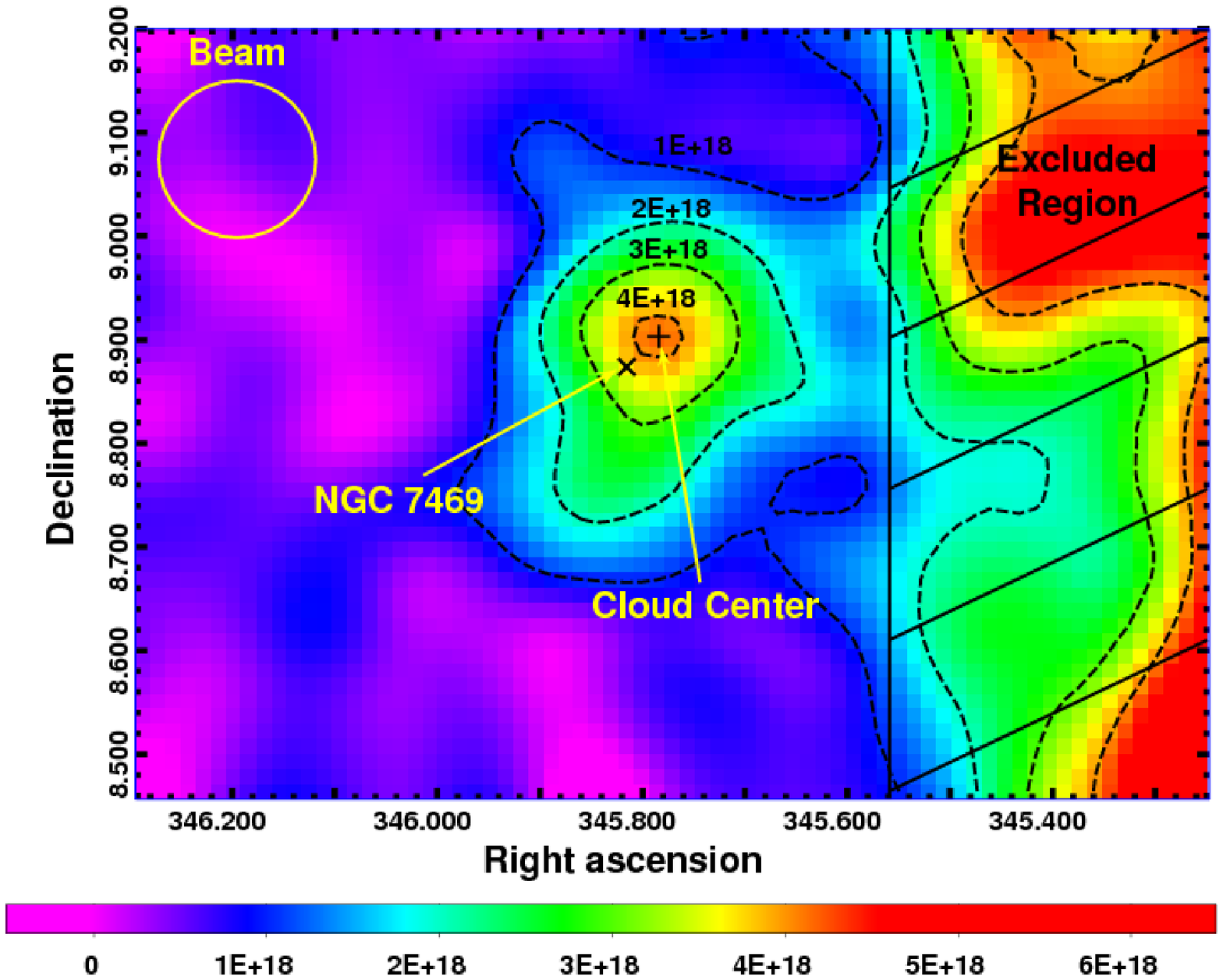}
\par\end{centering}

}\subfloat[Velocity (color scale in $\text{km\,\ s}^{-1}$)]{\begin{centering}
\includegraphics[width=0.38\paperwidth]{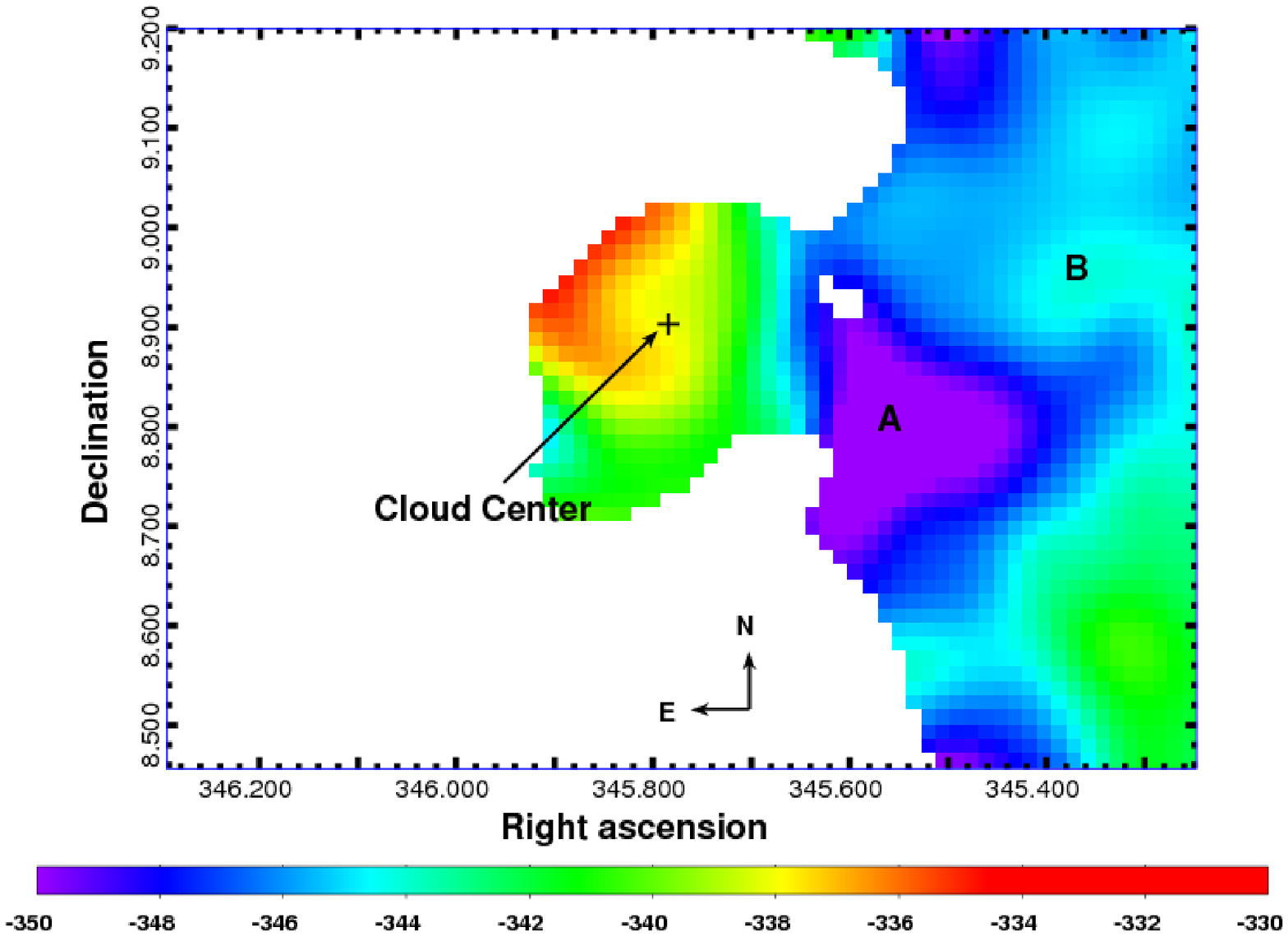}
\par\end{centering}

}
\par\end{centering}

\caption[Column density and velocity images of the cloud]{The \ion{H}{i} column density (a) and velocity (b) images of the
region containing the NGC~7469 sight line and the \ion{H}{i} cloud
that it passes through. The velocity range is $-360$~to~$-320\,\text{km}\,\text{s}^{-1}$
and the velocity image column density threshold is $2\times10^{18}\,\text{cm}^{-2}$.
The images show that, although the large cloud complex to the west
has a similar velocity range, the cloud is connected to it only by
diffuse bridging gas and is well isolated in velocity. The cloud complex
region was excluded from the circular path averaging as indicated.
\label{fig:ClumpMaps}}
\end{figure*}

In the data cube we unexpectedly found a roughly circularly projected,
mostly isolated \ion{H}{i} cloud in the MS, located only $\simeq2.5'$
from the direction of the well-studied Seyfert~1 background source
NGC~7469. Figure~\ref{fig:ClumpMaps} 
shows the \ion{H}{i} column density (zeroth moment) and velocity
(first moment, intensity-weighted) images of the cube over the velocity
range $-370$ to $-310\,\text{km}\,\text{s}^{-1}$. The cloud size
is $\approx1.6\times$beamwidth. After de-convolving the telescope
beam, the intrinsic cloud size is $\approx1.2\times$beamwidth. The
cloud is therefore just resolved. The \ion{H}{i} column density
at the cloud center is $\approx5\times10^{18}$ cm$^{-2}$ and although
it trails off to the north, east and south, it connects to the cloud
complex to the west. The velocity image shows that this connection
involves an abrupt $\simeq-10\,\text{km\,\ s}^{-1}$ velocity shift
from the $\simeq-340\,\text{km\,\ s}^{-1}$ gas at the cloud's apparent
western edge to the neighboring $\simeq-350\,\text{km\,\ s}^{-1}$
component labeled ``A'' of the western complex, which is mixed in
with a nearby $\simeq-345\,\text{km\,\ s}^{-1}$ component labeled
``B''. Note also that the cloud itself has a distinct $\simeq+5\,\text{km\,\ s}^{-1}$
gradient from west to east. The abrupt positive velocity shift from
the western complex along with the continued positive gradient across
the cloud suggests it may have broken off from the complex and is
becoming entrained in the surrounding ambient gas.

\begin{figure*}
\begin{centering}
\includegraphics[width=0.5\paperwidth]{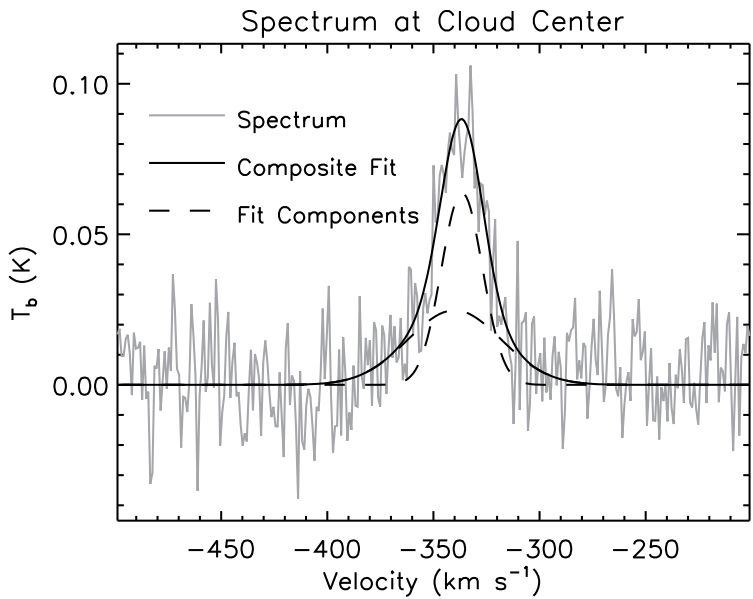} 
\par\end{centering}

\caption[Cloud center spectrum.]{\ion{H}{i} spectrum at the cloud center reveals two components,
a $\approx20\,\text{km}\,\text{s}^{-1}$ wide component at $\approx-340\,\text{km}\,\text{s}^{-1}$
on top of a weaker, $\approx60\,\text{km}\,\text{s}^{-1}$ wide component.
Individual Gaussian fits to each component are shown (dashed lines)
as well as the composite fit (solid line).\label{fig:HICentSpect}}
\end{figure*}

Figure~\ref{fig:HICentSpect} shows the \ion{H}{i} spectrum at
the center of the cloud. The emission clearly has a component at $\approx-340\,\text{km}\,\text{s}^{-1}$
with a $\text{FWHM}\approx20\,\text{km}\,\text{s}^{-1}$. However,
an even lower level, broad component lies at slightly more negative
LOS velocities as made evident by the two-component Gaussian fit overlaid
on the spectrum. The width and weakness of the wider envelope line
causes us to consider the possibility that it is an artifact of the
baseline removal process, but we reject that as there is independent
evidence of components associated with both the core and envelope
line velocities in the UV absorption spectra along the nearby NGC~7469
sight line reported in FSW10 (discussed in Section~\ref{sec:Discussion}).

To enhance sensitivity of our observations and characterize how properties
of the two components evolve outward from the cloud center, we introduce
the spatial averaging method and apply it to the data cube in Section~\ref{sub:SpaAveMeth:RawCloudProfile}.
This produces a profile of average column spectra versus projected
angular distance, $\rho$ from the center with increasing sensitivity,
well past its apparent periphery. In Section~\ref{sub:SpaAveMeth:LineModeling}
we purge the average profile of components that are unassociated with
the cloud. Finally, in Section~\ref{sub:3DCloudModel}, we develop
a simple parametric 3D spherical model of the cloud optimally matched
to the purged 2D projected profile to obtain a robust estimate of
the properties and structure of the cloud versus physical distance,
$r$ from its physical center outward.

\subsection{Spatial Averaging Method: Raw cloud Profile}

\label{sub:SpaAveMeth:RawCloudProfile}

To perform spatial averaging of \ion{H}{i} spectra we start with
the following assumptions: (i) the cloud is roughly circularly symmetric,
(ii) the cloud symmetry extends past its apparent boundaries, and
(iii) along each line of sight on a circular annulus from the cloud
center, \ion{H}{i} gas has similar properties. The small velocity
gradient of $\simeq+5\,\text{km\,\ s}^{-1}$ on the eastern side of
the clump (Figure~\ref{fig:ClumpMaps}(b), discussed above) shows
that the symmetry assumption is compromised, but the gradient is relatively
small with respect to the width of both cloud components identified
in the central spectrum of Figure~\ref{fig:HICentSpect}. Its effect
on the results of the averaging process that follows are considered
in Sections~\ref{sub:SpaAveMeth:LineModeling} and \ref{sub:3DCloudModel}.

\begin{figure*}
\begin{centering}
\includegraphics[width=8cm]{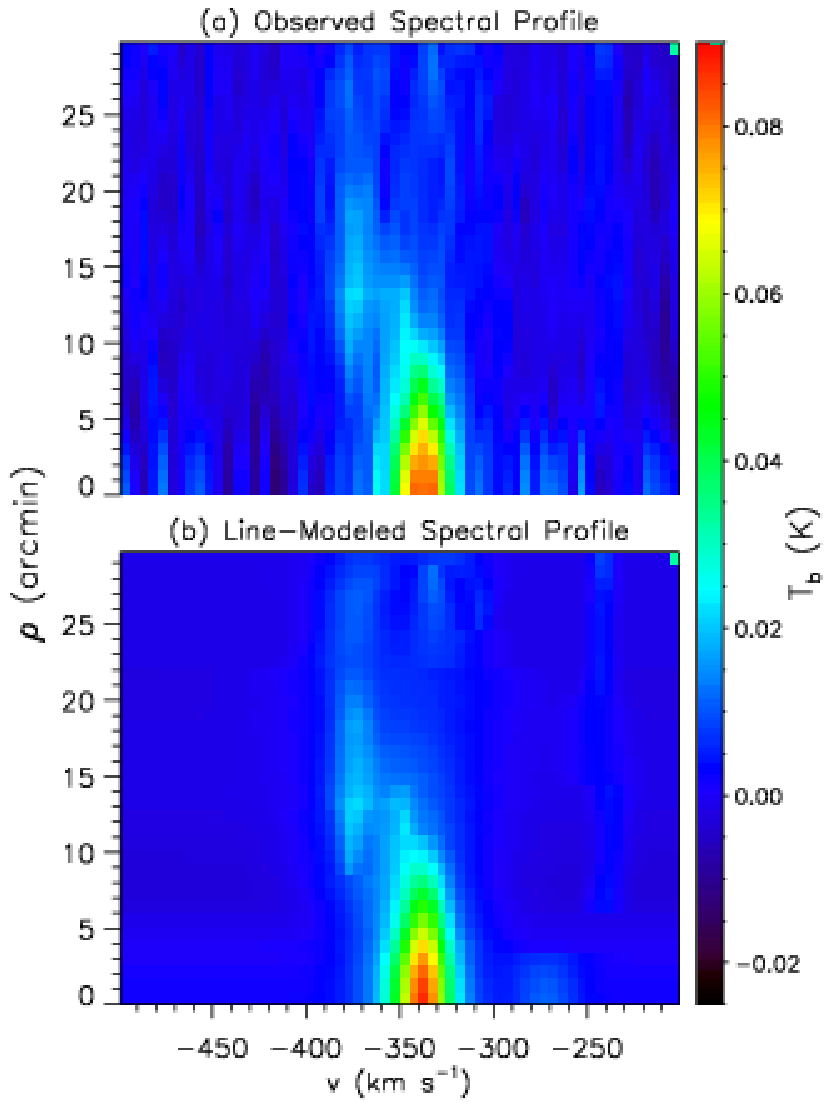} \includegraphics[width=8cm]{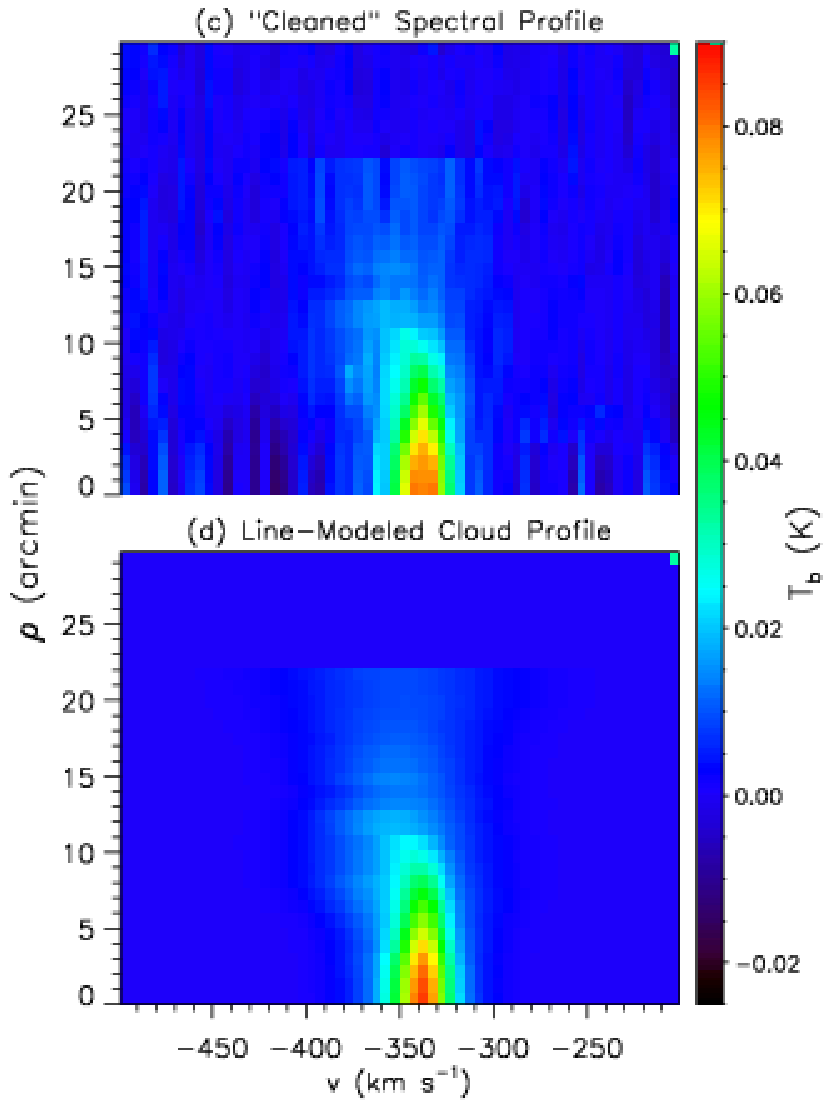} 
\par\end{centering}

\caption[Spatially averaged profiles. Line modeling phase]{Spatially averaged spectral profiles in the line modeling phase of
the analysis where the measured profile is ``cleaned'' of emission
not related to the cloud: (a) Observed spectral profile of the cloud
showing the intensity spectrum (horizontal axis) averaged along circular
paths of increasing distance from the center (vertical axis). (b)
Blended Gaussian emission line model of the entire profile. (c) Cleaned
spectral profile of the cloud where Gaussian emission line models
of all except the lines associated with the two components of the
cloud's central region are subtracted is used to constrain the optimally
matched 3D parametric cloud model in the next step. (d) Lines associated
with the two cloud components. For image display purposes only, profiles
have been smoothed from $0.966\,\text{km}\,\text{s}^{-1}$ to $4.83\,\text{km}\,\text{s}^{-1}$.
Note that the vertical striations in the noise are the consequence
of a noise correlation length of roughly four pixels due to interpolation.
\label{fig:SpectProfileLineMod}}
\end{figure*}

We start from the pixel with the highest \ion{H}{i} brightness
temperature (see Figure~\ref{fig:ClumpMaps}) and average \ion{H}{i}
spectra spatially along circular annuli of increasing projected radius
$\rho$. As a result, we can plot the averaged \ion{H}{i} spectra
as a function of projected angular distance from the cloud center
as shown in Figure~\ref{fig:SpectProfileLineMod}(a). The most important
advantage of this spatial averaging method is that it reduces the
noise without compromising angular resolution in the essential radial
dimension, which contains most of the spatial information in the nominally
circularly projected cloud. The noise improvement in the averaged
spectra increases with $\rho$ due to an increasing averaging path
length, thus compensating for decreasing emission from the cloud center.
Assuming infinite resolution and uncorrelated, zero mean noise at
each pixel, the measurement error for a circular cloud theoretically
decreases as $1/\sqrt{2\pi\rho}$. In reality, error improvement will
be limited to the extent that the noise is zero mean and uncorrelated
from pixel to pixel. The normal gridding process increases pixel-to-pixel
correlation, while residual structure from imperfect baseline modeling
violates both of these criteria and will generally be the limiting
factor in most applications of this method. While we apply the method
here to a roughly circular cloud, we note that the same method can
be applied on filamentary structures by assuming linear symmetry and
averaging along lines parallel to the filament main axis.

The result of the spatial averaging method is shown in Figure~\ref{fig:SpectProfileLineMod}(a).
 On the x-axis we plot the averaged \ion{H}{i} spectra at a given
projected distance $\rho$ from the cloud center, shown on the y-axis.
The first spectrum at $\rho=0\,'$ corresponds to the single \ion{H}{i}
spectrum through the cloud center (shown in Figure 3), and is dominated
by the main emission component centered at $v\simeq-340\,\text{km}\,\text{s}^{-1}$.
As $\rho$ increases, the primary emission component weakens and the
broader, underlying component strengthens as the sensitivity improves.
There is also a distinct blue-ward velocity gradient in the broad
component spectrum as $\rho$ increases. The distinct narrow and broad
emission components of the cloud are suggestive of a spherical structure
with a (kinematically) warm core and a warmer envelope: with increasing
$\rho$, the line of sight passes through less of the core and more
of the envelope and eventually, the envelope dominates through limb
brightening. We explore this further through modeling in Section~\ref{sub:3DCloudModel}.

In addition to emission at the main component velocities, there appear
to be one or more weak lines in the spectrum that are well separated
from the main emission (\emph{e.g.}, at $-250\,\text{km\,\ s}^{-1}$)
and therefore assumed to be unassociated with it. The evolving profile
also contains components at similar velocities, but unrelated to the
cloud's central components as the widening circle of averaging passes
through clumps of emission that are spatially separated on the sky.
These are also assumed to be unassociated. In the next step, these
unassociated components are purged from the projected profile.

\subsection{Spatial Averaging Method: Profile Line Modeling}

\label{sub:SpaAveMeth:LineModeling}

Using a semi-manual process aided by a graphically interactive program
written in IDL, all \ion{H}{i} spectra in the averaged profile
were modeled as multiple, blended Gaussian functions. The process
is as follows: Each \ion{H}{i} spectrum of the profile is viewed
in order of increasing $\rho$ and initial estimates of all potential
Gaussian components are provided by the user to a blended Gaussian
fitting routine. Each component of the resulting fit is then checked
and any with $\text{SNR}<2$ are removed with the remaining components
used as initial conditions for a second fit. At this point, since
Gaussian decomposition is not unique, the user views the fit and can
accept it or try again with new initial estimates.

The result of this modeling process is shown in Figure~\ref{fig:SpectProfileLineMod}(b).
While many unrelated lines pop in and out with increasing $\rho$,
there are two main Gaussian components that maintain continuity from
$\rho=0\,'$ outward and so are considered to be the components of
the cloud. All others are to be subtracted. It is important to stress
that the purpose of this line modeling step is \emph{not} to accurately
model all the emission in the profile but only to identify and subtract
components in the observed profile that are likely to be unrelated
to the cloud. The resultant ``cleaned'' profile, which contains
only the two observed primary components directly associated with
the cloud, is shown in \ref{fig:SpectProfileLineMod}(c). Note that
there is a prominent discontinuity in the cleaned profile which occurs
at the point where the broad component becomes undetectable as part
of the blended fit underneath two rising, but unrelated components
that dominate it. This is an artifact of imperfect line modeling.
Where the broad component is too weak to contribute its own line model
to the blended fit, the other components incorporate it into their
line models and then it gets subtracted. 

\begin{figure*}
\begin{centering}
\subfloat[Line Width]{\begin{centering}
\includegraphics[width=0.2\paperwidth]{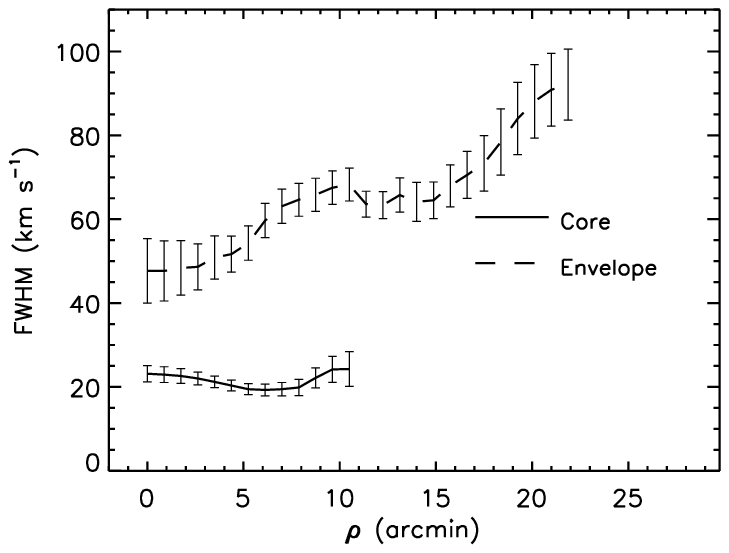}
\par\end{centering}

}\subfloat[Central velocity ]{\begin{centering}
\includegraphics[width=0.2\paperwidth]{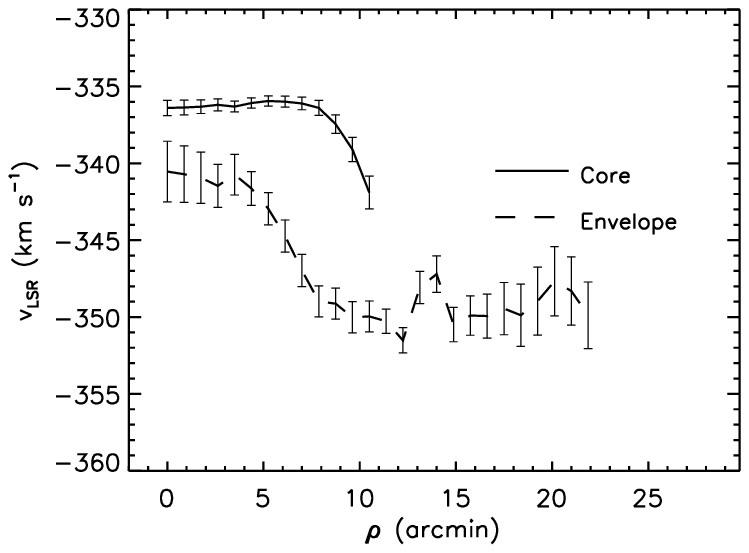}
\par\end{centering}

\centering{}}\subfloat[Column density]{\begin{centering}
\includegraphics[width=0.2\paperwidth]{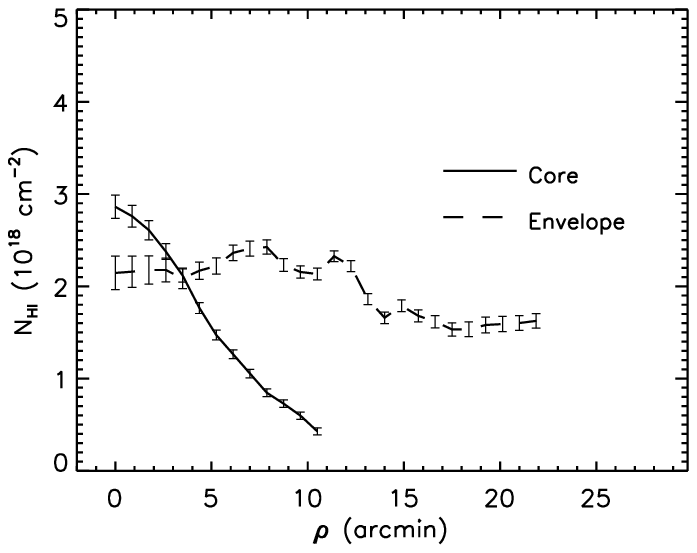}
\par\end{centering}

\centering{}}
\par\end{centering}

\caption[Cloud core and envelope line model parameter profiles]{Cloud core and envelope Gaussian component line model parameter profiles.
(a) Line Width (FWHM); (b) Velocity; and (c) Column Density. Error
bars reflect parameter errors due to noise only. Other effects such
as cross-coupling between components (see text) are not accounted
for. \label{fig:SpecProfileLineParms}}
\end{figure*}

Another limitation of the line modeling is cross-coupling, where a
stronger component affects the parameters of a weaker overlapping
component. These effects are not easily quantified and could dominate
any uncertainties due to noise alone. Imperfections notwithstanding,
the lines associated with the two components of the cloud are modeled
as part of the cleaning process and provide a rough model of the cleaned
profile as shown in Figure~\ref{fig:SpectProfileLineMod}(d). Figure~\ref{fig:SpecProfileLineParms}
shows how the modeled line width, central velocity, and \ion{H}{i}
column density parameters of these two components evolve with $\rho$.
.
 Figure~\ref{fig:SpecProfileLineParms}(a) shows that the two components
have distinctly different line widths. The narrow core component has
a relatively constant $\text{FWHM}\simeq20\,\text{km\,\ s}^{-1}$
while the broader envelope has an apparently large outward gradient
from $\simeq50\,\text{km\,\ s}^{-1}$ at cloud center to $\simeq80\,\text{km\,\ s}^{-1}$
at $\rho\sim20\,'$ where the line falls below threshold. Although
the gradient may be exaggerated by parameter cross-coupling with the
core component at low $\rho$ and with subtracted components at higher
$\rho$, it could be significant and is physically consistent with
turbulent and/or thermal broadening as it approaches the hot ambient
Halo medium.

The central velocity profiles of Figure~\ref{fig:SpecProfileLineParms}(b)
illustrate the parameter coupling concerns mentioned previously. At
small $\rho$ the stronger core component may ``pull'' the velocity
of the weaker envelope, while the roles are reversed at $\rho\sim9\,'$,
where the core weakens and its velocity drops sharply toward the envelope's.
The sharp drop in core velocity and the step change in envelope velocity
are probably not entirely real. The core velocity is likely constant
for the most part at $\simeq-336\,\text{km\,\ s}^{-1}$ with the sharp
drop clearly a modeling artifact since this is not possible given
the beam smoothing. The envelope's apparent negative velocity gradient
may be exaggerated by ``pulling'' from the core at low $\rho$ but
is large enough that it could be at least partially real, approaching
$\simeq-350\,\text{km\,\ s}^{-1}$ at the periphery. 

Gradient or not, there \emph{is} a difference in the central velocity
between the two components of between 4 and $14\,\text{km\,\ s}^{-1}$
with the core component clearly \emph{lagging} the envelope component
along the LOS. This is very interesting because the large negative
velocity of this cloud indicates infall and a cometary morphology
with the core \emph{leading} the envelope is expected. Although the
morphology may be hidden in projection or in the noise of Figure~\ref{fig:ClumpMaps},
the core in the spatially integrated profile is clearly lagging in
velocity. This is also apparent in Figure~\ref{fig:HICentSpect}
(the central cloud spectrum) indicating that this is not an artifact
of the spatial averaging process. \citet{bruns01} similarly modeled
the line profiles of HVC125+41-207 along simple perpendicular cross
sections (\emph{i.e.}, no spatial averaging) and demonstrated that
its projected on-sky head-tail morphology had a cold core leading
a warmer envelope. Similarly, many other HVCs with a head-tail structure
have a lagging tail, which is usually interpreted as belonging to
gas stripped and slowed down through interactions with the surrounding
medium. This interesting, apparently anomalous velocity structure
of our cloud is addressed in Section~\ref{sec:Discussion}. Some
of the gradient in line width and central velocity of Figure~\ref{fig:SpecProfileLineParms}(a)
and (b) could be caused by the velocity gradient asymmetry toward
the eastern side of the clump. We discuss this in Section~\ref{sub:3DCloudModel}.

Figure~\ref{fig:SpecProfileLineParms}(c) shows the well-behaved
\ion{H}{i} column density of the two components. The core component
is roughly Gaussian with FWHM~$\simeq10\,'$, only slightly larger
than the $9.1\,'$ beamwidth. This reflects the barely resolved core.
The envelope component has an almost constant column density of $\simeq2\times10^{18}\,\text{cm}^{-2}$.
It is worth emphasizing that the broad envelope component which has
a significant \ion{H}{i} column density even at large $\rho$,
constitutes a large portion of the cloud's total neutral mass and
probably an even larger portion of its total mass since it is likely
more highly ionized than the core. In Section~\ref{sub:3DCloudModel}
we quantify this in the context of the 3D cloud model developed there.

\begin{figure*}
\begin{centering}
\includegraphics[width=8cm]{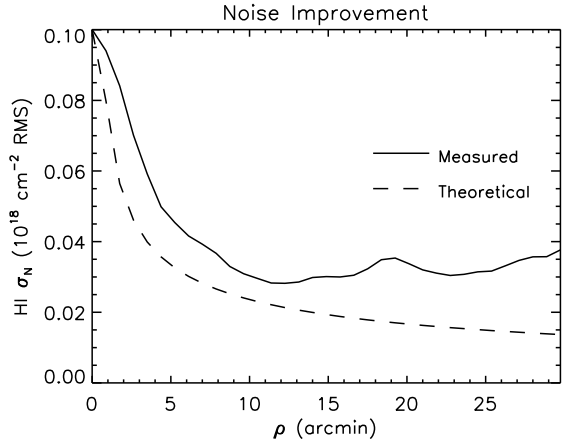} 
\par\end{centering}

\caption[RMS column density noise]{The measured RMS noise of the \ion{H}{i} column density for a
$15\,\text{km\,\ s}^{-1}$ FWHM line (solid line) versus the projected
distance from the cloud center. Also shown is the theoretical error
improvement (dashed line) assuming zero mean, uncorrelated noise and
infinite resolution. Measured noise does not improve as quickly at
first partially because the effects of finite resolution are more
important for short averaging paths. Note that close to $\rho\sim11\,'$
measured noise approaches the theoretical value of $2.4\times10^{16}\,\text{cm}^{-2}$,
and then increases due to path truncation from exclusion of the western
region of the cube. This is a very deep measurement and maintains
error in the periphery below $4.0\times10^{16}\,\text{cm}^{-2}$.\label{fig:RefSensitivty}}
\end{figure*}

As a by-product of the component fitting process the residual noise
of each averaged \ion{H}{i} spectrum is measured after subtracting
the line fit. This is plotted in Figure~\ref{fig:RefSensitivty}
 as a function of projected angular distance $\rho$ from the cloud
center. This figure clearly demonstrates the extremely high sensitivity
of this spatial averaging method, where $\sigma_{N}$ reaches below
$4\times10^{16}\,\text{cm}^{-2}$ at about $10\,'$ from the cloud
center. Also plotted is the theoretical noise as discussed in Section~\ref{sub:SpaAveMeth:LineModeling},
which also incorporates the noise correlation effects of interpolation.
Measured noise does not improve as rapidly as it could, probably because
of limited resolution and residual baseline artifacts at low $\rho$
and the correlation introduced by the gridding process as mentioned
before. Note that at the beamwidth, $\rho\approx9\,'$, correlation
effects are minimal and the noise improvement approaches the theoretical
limit. However, at $\rho>12\,'$, the averaging path begins to be
truncated by the excluded region and then by the borders of the cube
and so the noise flattens and begins to increase.

\subsection{3D Cloud Model Fitting}

\label{sub:3DCloudModel}

We now use the two-dimensional projected profile of Figure~\ref{fig:SpectProfileLineMod}(c)
to investigate a simple three-dimensional, core-envelope cloud model
supporting our observational claims regarding the cloud structure.
Since the cloud is not spatially resolved well in our measurements,
subtle spatial structure in the model is not appropriate, but since
spectral resolution is very high, subtle spectral structure can be
incorporated. We define a spherically symmetric cloud model with three
spatial and one velocity dimension as:

\begin{equation}
\begin{split}n & _{HI}(r,v)=\\
 & \begin{cases}
\frac{n_{c0}}{\sqrt{2\pi}\sigma_{c}(r)}e^{-\frac{(v-v_{c0})^{2}}{2\sigma_{c}^{2}(r)}} & \text{, \ensuremath{r<r_{c}}}\\
\frac{n_{e0}}{\sqrt{2\pi}\sigma_{e}(r)}e^{-\frac{(v-v_{e}(r))^{2}}{2\sigma_{e}^{2}(r)}} & \text{, \ensuremath{r_{c}\leq r\leq r_{c}+r_{e}}}\\
0 & \text{, \ensuremath{r>r_{c}+r_{e}}}
\end{cases}
\end{split}
\label{eq:EmissionModel-1}
\end{equation}
 where 
\begin{align}
\sigma_{c}(r) & =\sigma_{vc0}+m_{\sigma c}r,\\
\sigma_{e}(r) & =\sigma_{ve0}+m_{\sigma e}r,\\
v_{e}(r) & =v_{e0}+m_{ve}r,\\
r & =\sqrt{x^{2}+y^{2}+z^{2}}
\end{align}
 and cloud spatial coordinates $x,y,z$ are defined so that the $z$-axis
is along the LOS, $x$ is east to west on the sky, while $y$ is north
to south. The cloud consists of a core with a constant \ion{H}{i}
volume density $n_{c0}$ and has a radius $r_{c}$. Beyond the core
is the cloud envelope with a constant density $n_{e0}$ which extends
up to a radius of $r_{e}$. At the cloud center, $r=0$, the core
and envelope have a LOS velocity $v_{c0}$ and $v_{e0}$, respectively.
The spectra of the core and envelope are Gaussian functions centered
at $v_{c0}$ and $v_{e0}$, respectively. The velocity dispersion
$\sigma_{v}$ of the core is allowed to linearly vary with $r$, but
the central velocity is held constant since the observed column profile
(Figure~\ref{fig:SpecProfileLineParms}) shows little variation.
Significant gradient in the velocity and dispersion of the envelope
is suggested by the profile, so both are allowed to vary linearly
with $r$.

Note that this model incorporates a radially dependent velocity scalar
field that is restricted to the LOS component of the vector field.
As such, the model, although very sensitive to the directly measured
LOS motions associated with a cometary feature seen in projection,
is insensitive to radial motion from the cloud's 3D center. So, expansion
or contraction effects cannot be inferred from the model. Parametric
variables used in this model are defined in Table~\ref{tab:ModCloudParms}.
\begin{deluxetable*}{clc}  % <--- column justification (center/left/right)
\tablecolumns{3}
\tablewidth{0pt}
\tabletypesize{\small}
\tablecaption{Core plus Envelope Model Opimized Parameters\label{tab:ModCloudParms}}
\tablehead{   % column headings
  \colhead{Parameter} &
  \colhead{Description\tablenotemark{a}} &
  \colhead{Value} 
}
\startdata
{$r_{c}$} & {Core normalized radius (arcmin)} & {$7.32\pm0.37$}\\
{$n_{c0}$} & {Core number density at center ($\frac{70\,\text{kpc}}{D}\text{cm}^{-3}$)} & {$(5.27\pm0.41)\times10^{-3}$}\\
{$v_{c0}$} & {Core velocity (km~s$^{-1}$)} & {$-336.3\pm0.2$}\\
{$\sigma_{vc0}$} & {Core velocity dispersion at center (km~s$^{-1}$)} & {$8.80\pm2.94$}\\
{$m_{\sigma c}$} & {Core velocity dispersion slope ($\frac{\text{km\,\ s}^{-1}}{\text{arcmin}}$)} & {$0.20\pm0.61$}\\
{$r_{e}$} & {Envelope normalized thickness (arcmin)} & {$18.51\pm0.41$}\\
{$n_{e0}$} & {Envelope density ($\frac{70\,\text{kpc}}{D}\text{cm}^{-3}$)} & {$(8.83\pm0.26)\times10^{-4}$}\\
{$v_{e0}$} & {Envelope initial velocity (km~s$^{-1}$)} & {$-349.6\pm0.4$}\\
{$m_{ve}$} & {Envelope velocity slope ($\frac{\text{km\,\ s}^{-1}}{\text{arcmin}}$)} & {$-0.441\pm0.264$}\\
{$\sigma_{ve0}$} & {Envelope initial velocity dispersion (km~s$^{-1}$)} & {$25.8\pm3.0$}\\
{$m_{\sigma e}$} & {Envelope velocity dispersion slope ($\frac{\text{km\,\ s}^{-1}}{\text{arcmin}}$)} & {$0.157\pm0.253$}\\
\enddata
\tablenotetext{a}{Scalable units use $D$, the LOS distance in pc.}
\end{deluxetable*}

In practice, the function of Equation~\eqref{eq:EmissionModel-1}
was used to generate a model grid with three spatial dimensions plus
one for LOS velocity. The spatial grid pixel spacing was the same
as that of the observed radial profile. The grid encompassed the entire
spherical cloud which spans twice the profile's radial extent in each
dimension, but was extended in the $x$ and $y$ dimensions by two
beamwidths at zero density to allow for beam smoothing. The spatial
averaging method was applied to the 3D model by summing the spectra
in the $z$ direction at each $(x,y)$ pixel to obtain a spectral
profile to be compared with the observed cloud profile. The weighting
used was the inverse of the measured profile variance at each pixel.
Prior to comparing simulated and observed profiles, the model array
was processed to simulate observing effects. This involved approximate
beam smoothing with a 2D Gaussian having FWHM equal to the GBT beamwidth,
truncated at $\pm\text{FWHM}$. An equivalent averaging of spectra
along circular paths was then applied as was done in the case of the
observed data cube.

We used the AMOEBA algorithm \citep{nelder65} to find the model parameters
that minimize the weighted least mean-squared error of the 3D model's
simulated observed profile. AMOEBA finds the local minimum in the
parameter space that is nearest to the initial guess. All of the parameters
in this model relating to the velocity spectrum shape are easily estimated
from and well constrained by the observed profile data. They are also
relatively independent of the core and envelope sizes. For these parameters,
the error surface should be well behaved with a single minimum. The
core and envelope sizes and densities are highly inter dependent,
and the error surface would have a very broad minimum given that the
beam smoothing is comparable to the profile's scale width. Also, there
may very well be multiple minima. Because of these concerns, a sweep
of the initial guesses for these two size parameters was performed
in order to find the lowest local optimum over a range of values.
For each combination of initial core and envelope size guesses, approximately
self-consistent initial density values were derived from the observed
column density at $\rho=0$ and $\rho\approx r_{c}$, respectively.
The optimization was run at these combinations and the solution with
the minimum error at optimum was chosen. Optimized parameter values
are shown in Table~\ref{tab:ModCloudParms} along with $1\sigma$
error estimates. The parameter error vector $\mathbf{\mathbf{\Delta}p}=(\Delta p_{n})$
was estimated by taking the square root of the diagonal elements of
the covariance matrix $[\,\mathbf{J^{T}W\, J}\,]^{-1}$. Here, $\mathbf{W}=[1/\sigma_{m}^{2}]$
is the diagonal weighting matrix used in the optimization, $\sigma_{m}$
are measured error vector elements (flattened matrix of rms noise
residuals from line modeling), and $\mathbf{J}=[J_{mn}]=[\frac{\partial y_{m}}{\partial p_{n}}]$
is the Jacobian matrix at the optimum solution (determined numerically).

\begin{figure*}
\begin{centering}
\includegraphics[width=8cm]{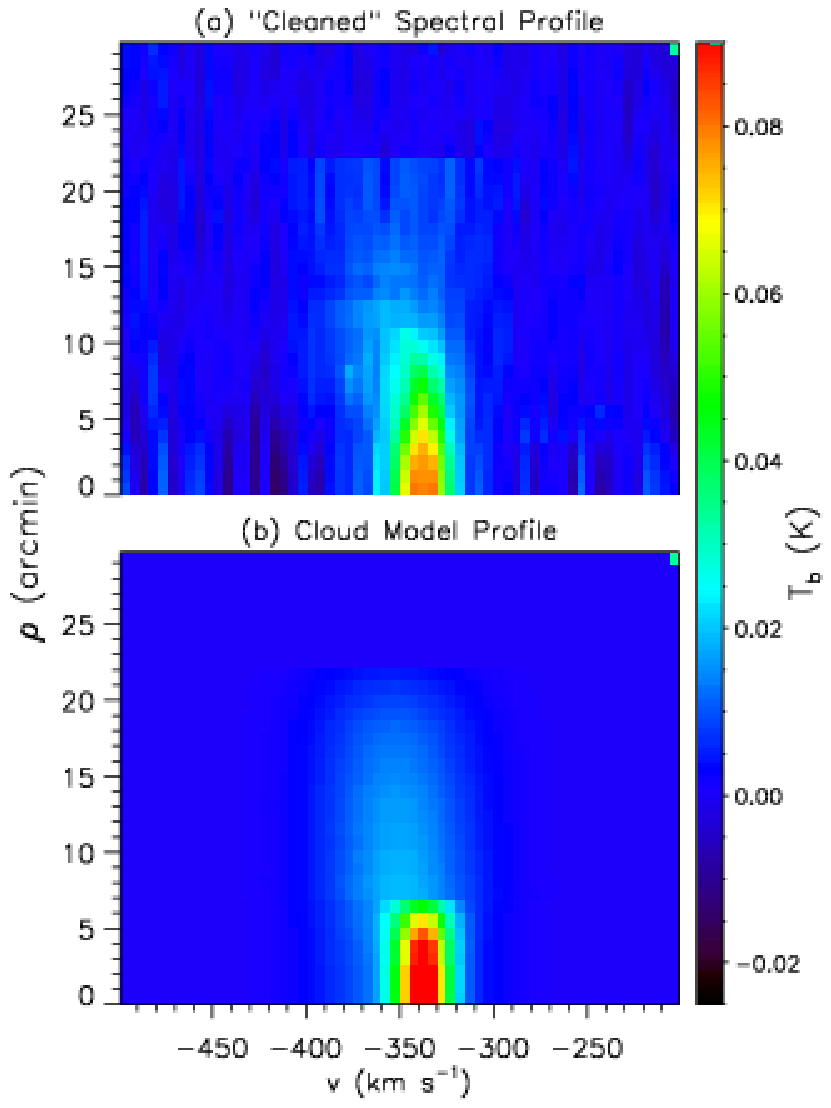} \includegraphics[width=8cm]{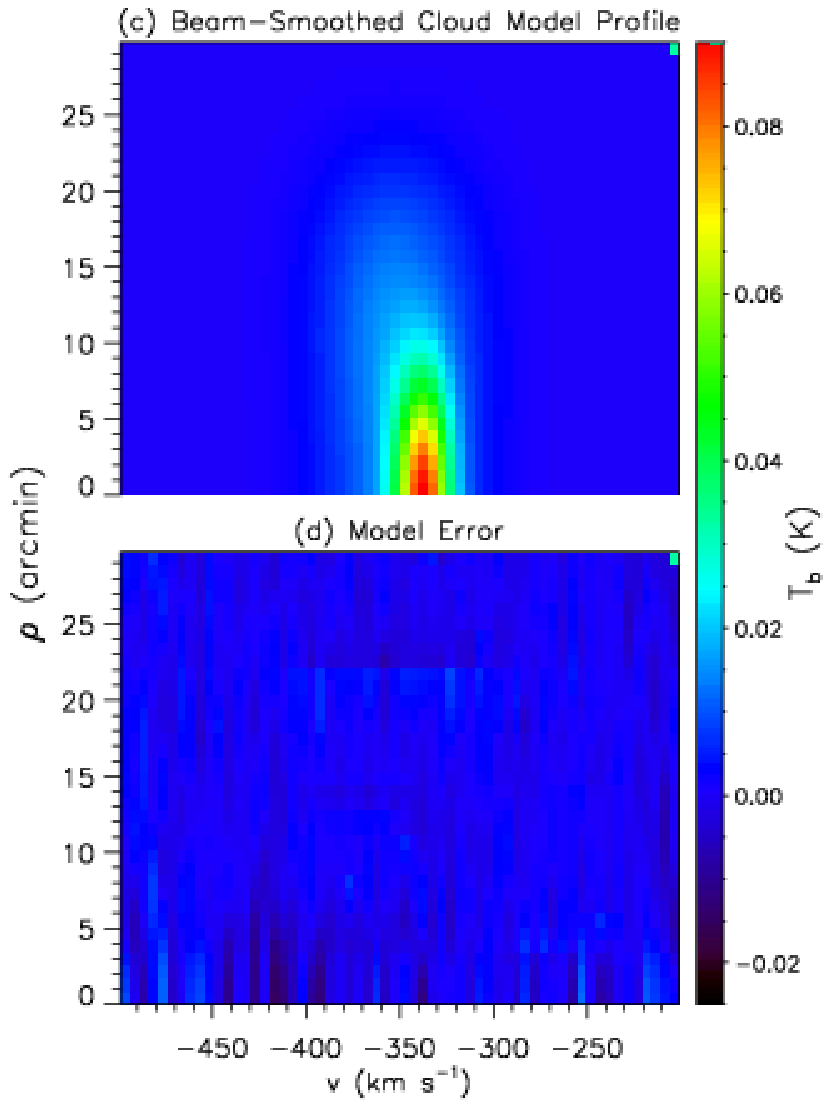} 
\par\end{centering}

\caption[Cloud modeling results]{Cloud modeling results: (a) Cleaned spatially averaged spectral profile
purged of non-cloud components. (b) The projected spectral profile
of the two-component 3D cloud model without beam smoothing. (c) The
projected model profile with beam smoothing. (d) The model profile
residual error, or the difference between the cleaned profile and
the beam-smoothed model profile. For image display purposes only,
profiles have been smoothed from $0.966\,\text{km\,\ s}^{-1}$ to
$4.83\,\text{km\,\ s}^{-1}$.\label{fig:SpecProfileCloudModel}}
\end{figure*}

Figure~\ref{fig:SpecProfileCloudModel} 
 shows the cloud modeling results. Figure~\ref{fig:SpecProfileCloudModel}(a)
is the spatially averaged observed projected spectral profile purged
of all but the main emission component (same as in Figure~\ref{fig:SpectProfileLineMod}(c)),
used as the optimization target for the AMOEBA algorithm. Figure~\ref{fig:SpecProfileCloudModel}(b)
is the projected spectral profile of the optimized 3D cloud model
without beam smoothing effects, Figure~\ref{fig:SpecProfileCloudModel}(c)
is the same as (b) with beam smoothing, which is used to compare with
\emph{i.e.}, subtract from, (a) in the optimization process. Figure~\ref{fig:SpecProfileCloudModel}(d)
shows the residual from this subtraction. The residual structure at
$\rho$ from $\simeq20\text{ to }24\,'$ is a consequence of the optimization
``smoothing'' out the discontinuity in the cleaned envelope emission
at the points where it meets line detection threshold. The low residual
levels in the image show that our simple core-envelope 3D cloud model,
when similarly projected onto a 2D spectral profile, clearly compares
well with the observed data. This supports our interpretation of narrow
and broad velocity components tracing internal structure of a single
\ion{H}{i} cloud.

The quality of the fit for the model is quantified by comparing the
RMS of this modeling residual, $\sigma_{model}=6.17\,\text{mK}$ (calculated
between $-450$ and $-250\,\text{km\,\ s}^{-1}$, the range of emission),
to the RMS noise of the observed profile, $\sigma_{noise}=6.20\,\text{mK}$
measured in the previous line modeling step. This represents a $\sqrt{{|\sigma_{model}^{2}-\sigma_{noise}^{2}|}}=0.7\,\text{mK}$
RMS modeling error, only $11\%$ of the noise. The simplicity of this
model and the low residual error strongly suggest that this is a reasonable
representation of the observed data. We therefore proceed with further
data analysis under the core-envelope cloud model.

\begin{figure*}
\begin{centering}
\subfloat[Component density]{\begin{centering}
\includegraphics[width=5cm]{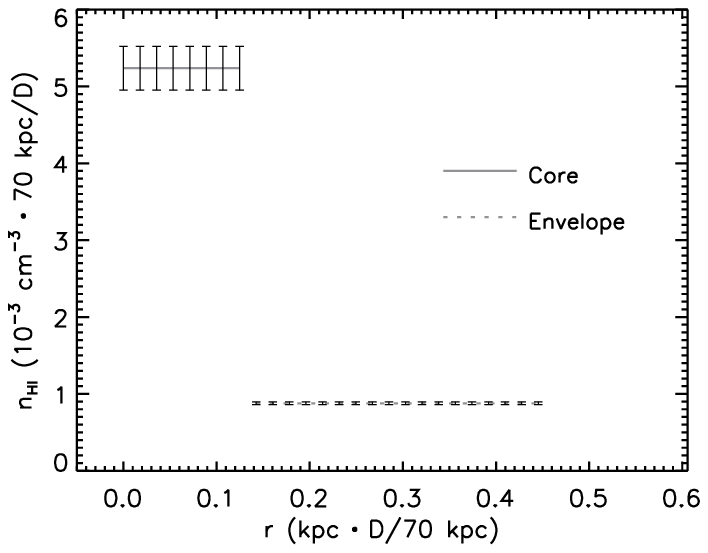}
\par\end{centering}

}\subfloat[Component velocity ]{\begin{centering}
\includegraphics[width=5cm]{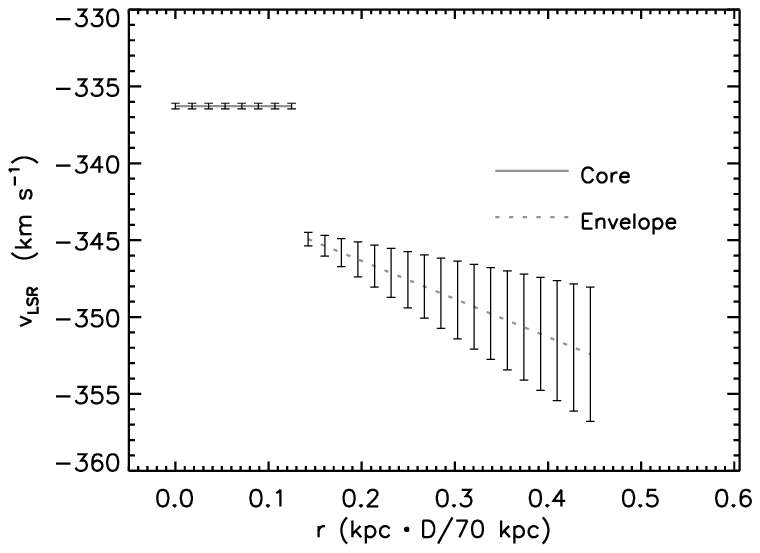}
\par\end{centering}

} \subfloat[Component line width ]{\begin{centering}
\includegraphics[width=5cm]{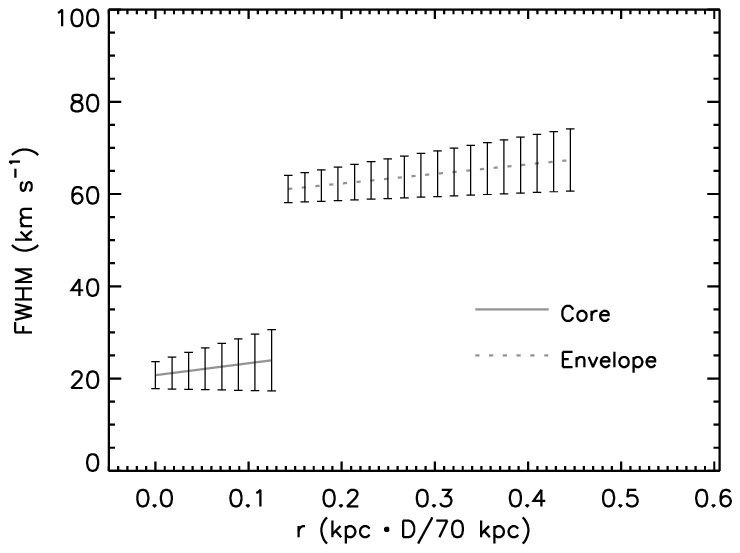}
\par\end{centering}

}
\par\end{centering}

\caption[Characteristics of two component cloud model.]{Characteristics of the optimized two-component core-plus-envelope
spherically symmetric cloud model: (a) Stepped uniform density profile
as imposed by the model. (b) The velocity profile shows an abrupt
blue-ward shift at the core-envelope transition followed by a moderate
blue-ward gradient in the envelope, suggesting an outward motion of
the core with respect to the envelope, but both are only $\sim15\%$
of the envelope line width and therefore, not very significant. (c)
Line width profile with the core consistent with a WNM phase, and
the envelope $\sim9\times$ kinematically warmer. The line width gradients
are not significant given the errors, but consistent with warming
and/or increasing turbulence toward the periphery. These small velocity
and line width gradients may also include the effects of a small asymmetry
in the velocity structure of the cloud (see text). \label{fig:CloudModelParms}}
\end{figure*}

The model parameters are summarized in Figure~\ref{fig:CloudModelParms},
where values of \ion{H}{i} volume density, central line velocity
and line width for the two components are plotted as a function of
radial distance, $r$ from the cloud center.
 The best-fit values for the \ion{H}{i} volume density of the cloud
core and envelope are $5.3\times10^{-3}$ and $8.8\times10^{-4}$
cm$^{-3}$, respectively, scaled to a reference distance of 70 kpc.
For comparison, using an independent method \citet{stani02} estimated
the \ion{H}{i} volume density for several MS clouds in this part
of the MS at $\sim5\,\text{cm}^{-2}$ assuming a distance of 60 kpc.
The density of the modeled core scaled to this distance translates
to $6.1\times10^{-3}$ cm$^{-3}$, which is $\sim\nicefrac{1}{8}$
of that estimate, but the former is based on clumps with $>5$ times
the central column density so a higher neutral fraction is to be expected
due to more effective self-shielding. The LOS velocity of the model
core is constant, as constrained by the model, and the nominal velocity
of the envelope gradually decreases from $-344$ to $-351\,\text{km\,\ s}^{-1}$.
The error bars indicate considerable uncertainty in this gradient
which is only $\simeq10\%$ of the line width.

As mentioned previously, the small observed asymmetry in the observed
velocity field in Figure~\ref{fig:ClumpMaps}(b) could contribute
to the velocity centroid and width gradients observed in Figure~\ref{fig:SpecProfileLineParms}
(and modeled in Figure~\ref{fig:CloudModelParms}). We can estimate
the magnitude of this effect by quantifying the observed velocity
gradient asymmetry in the core, assume that it continues into the
envelope, and then estimate its affect on the model's averaged gradients.
From Figure~\ref{fig:ClumpMaps}(b), $\simeq+5\,\text{km\,\ s}^{-1}$
gradient occurs over approximately one beamwidth, or $\approx10'$
resulting in a velocity gradient of $\approx+0.5\,\text{km\,\ s}^{-1}\,\text{arcmin}^{-1}$.
Since the asymmetry is present for less than one-half of the averaging
path, the combined effect is about one-half the measured gradient,
or $\approx+0.25\,\text{km\,\ s}^{-1}\,\text{arcmin}^{-1}$. Referring
to Table~\ref{tab:ModCloudParms}, note that this could account for
most of the observed core line width gradient $m_{\sigma c}$ which
is dominated by the uncertainty due to noise anyway, so the effect
of the asymmetry on the core line width gradient is insignificant.
The model did not include a core velocity gradient because there appeared
to be little variation, but this positive gradient asymmetry may have
compensated for a roughly equal and opposite gradient amounting to
only $\approx2\,\text{km\,\ s}^{-1}$ across the $\approx7\,'$ radius
of the core. We also note that the negative envelope line width gradient
$m_{\sigma e}$ could be partially compensated by the asymmetry, but
the $\approx+0.25\,\text{km\,\ s}^{-1}\,\text{arcmin}^{-1}$ effect
is comparable to the error due to noise and so not of significant
concern.

Table~\ref{tab:ModCloudProp} summarizes the global properties of
the model cloud in more physically relevant terms useful for discussion
in Section~\ref{sec:Discussion}. Those properties that are distance-dependent
are scaled for a reference distance of 70~kpc LOS. The units indicate
how the quantities scale at other distances. It is particularly noteworthy
that while the envelope has about a factor of two lower \ion{H}{i}
column density, it is quite extended and is seven times more massive
than the core.
\begin{deluxetable*}{clcc}  % <--- column justification (center/left/right)
\tablecolumns{3}
\tablewidth{0pt}
\tabletypesize{\small}
\tablecaption{Model Cloud Properties\label{tab:ModCloudProp}}
\tablehead{   % column headings
  \colhead{Parameter} &
  \colhead{Description\tablenotemark{a}} &
  \colhead{Value}}
\startdata
{$[l,b]$} & {Galactic coordinates ($\deg$)} & {$[83,-45]$}\\
{$\theta_{lg}$} & {LOS to Galactocentric angle, $D>50$~kpc ($\deg$)} & {$<9.4$}\\
{$r_{c}$} & {Core radius ($\frac{D}{70\,\text{kpc}}\text{pc}$)} & {$147$}\\
{$\bar{n}_{c}$} & {Mean Core neutral density ($\frac{70\,\text{kpc}}{D}\text{cm}^{-3}$)} & {$5.3\times10^{-3}$}\\
{$\bar{\sigma}_{vc}$} & {Mean Core velocity dispersion (km~s$^{-1}$)} & {$10$}\\
{$\bar{v}_{c}$} & {Mean Core LOS velocity (km~s$^{-1}$))} & {$-336$}\\
{$N_{HIc}$} & {Central Core \ion{H}{i} column density ($\text{cm}^{-2}$)} & {$4.8\times10^{18}$}\\
{$M_{HIc}$} & {Core \ion{H}{i} mass (($\frac{D}{70\,\text{kpc}})^2M_{\odot}$)} & {$2.5\times10^3$}\\
{$r_{e}$} & {Envelope thickness ($\frac{D}{70\,\text{kpc}}\text{pc}$)} & {$380$}\\
{$\bar{n}_{e}$} & {Mean Envelope neutral density ($\frac{70\,\text{kpc}}{D}\text{cm}^{-3}$)} & {$8.9\times10^{-4}$}\\
{$\bar{\sigma}_{ve}$} & {Mean Envelope velocity dispersion (km~s$^{-1}$)} & {$29$}\\
{$\bar{v}_{e}$} & {Mean Envelope LOS velocity (km~s$^{-1}$))} & {$-358$}\\
{$N_{HIe}$} & {Central Envelope \ion{H}{i} column density ($\text{cm}^{-2}$)} & {$2.1\times10^{18}$}\\
{$M_{HIe}$} & {Envelope \ion{H}{i} mass (($\frac{D}{70\,\text{kpc}})^2M_{\odot}$)} & {$1.8\times10^4$}\\
{$M_{HI}$} & {Total \ion{H}{i} mass (($\frac{D}{70\,\text{kpc}})^2M_{\odot}$)} & {$2.1\times10^4$}\\
{$M_{Hmin}$} &
  Minimum total H mass{\tablenotemark{b} (($\frac{D}{70\,\text{kpc}})^2M_{\odot}$)}
  & {$4.1\times10^5$}\\
\enddata
\tablenotetext{a}{Scalable units use $D$, the LOS distance in pc.}\tablenotetext{b}{Assumes $\text{\ion{H}{ii}/\ion{H}{i}}>19$ \citep{fox10}}
\end{deluxetable*}

\section{Physical Interpretation of the Model}

\label{sec:Discussion}

Using the highly sensitive analytical technique based on a simple
3D parametric model described in Section~\ref{sec:Results}, we have
obtained observational evidence of the cloud having a distinct neutral
core with a WNM-like velocity line width, surrounded by an envelope
with a much wider line width, described by the model parameters of
Figure~\ref{fig:CloudModelParms}. We now delve into physical interpretations
of the observed spatial and kinematic structure.

The most interesting and puzzling aspect of our results so far is
the negative velocity offset of the envelope with respect to the core
which ranges from $-5\,\text{km\,\ s}^{-1}$ at the center and gets
larger further out to $-10\,\text{km\,\ s}^{-1}$ implying that in
this infalling structure, the envelope, presumably consisting of gas
components stripped from the core and entrained by the relatively
static Halo, is falling faster, \emph{leading} the core. We would
expect a typical cometary feature to develop with entrained, envelope
gas trailing behind the core with a \emph{lagging} velocity as found
by \citet{bruns01} in a structural analysis of HVC125+41\textendash{}207
previously mentioned in Section~\ref{sub:SpaAveMeth:LineModeling}.
Their analysis of highly resolved observations of the HVC revealed
WNM-like gas in a cometary envelope surrounding a well-resolved CNM-like
core. In contrast, our cloud's barely resolved WNM core corresponds
to their entire cloud and our envelope corresponds to possible boundary
gas surrounding their HVC, well below their detection threshold. Still,
we might expect an analogous structure in the boundary gas, in transition
to the Halo. Our cloud shows no clear projected on-sky cometary feature,
but given the high LOS velocity it would probably be hidden in full
projection behind the cloud. However, this orientation would directly
show the lagging envelope velocity signature of a cometary feature
yet we see it clearly leading.

The most likely explanation for the leading envelope is suggested
by the velocity structure of Figure~\ref{fig:ClumpMaps}(b) which,
as discussed in Section~\ref{sec:Results}, indicates the cloud may
be a large clump of entrained gas originating from the western complex
and now lagging it. The detected envelope velocity range of $\simeq-350\,\text{km\,\ s}^{-1}$
to $\simeq-345\,\text{km\,\ s}^{-1}$ corresponds closely to the velocity
range of the two nearby components (A and B) of the western complex.
The detected envelope therefore, may be an extended, diffuse structure
that both the cloud and the western complex are embedded in. Although
of low \ion{H}{i} column density, $\sim2\times10^{18}$ cm$^{-2}$,
it is quite extensive and may contain a significant amount of \ion{H}{i}.

However, our spatial averaging method provides further constraints
on the nature of the diffuse envelope. Since the region excluded in
the spatial averaging includes most of the extended structure and
bridging gas to the west past $\simeq14\,'$ from the cloud center,
our 3D cloud model is not influenced by this western region. We can
therefore assume that although the projected circular symmetry assumption
of the averaging method is weaker, the core-envelope cloud model (Section
3.3) reasonably represents properties of the envelope gas in the vicinity
of the cloud, suggesting a physical association between the cloud
core and its immediate envelope. Therefore, we proceed with interpretation
under the assumption that the core is embedded in the surrounding
envelope, but that the envelope is part of a larger structure in the
local hierarchy closely associated with the core.

In addition to our \ion{H}{i} observations, the UV spectroscopy
of NGC~7469 by FSW10 provides rich complementary information about
the cloud. Numerous absorption lines have been detected from both
low and high ionization species at MS velocities. FSW10 concluded
that neither photoionization nor single-temperature collisional ionization
can explain the observed column densities of high ions. This suggested
the existence of highly multi-phase plasma, with a cooler region traced
by \ion{Si}{iv} (peaking at $10^{4.8}\,\text{K}$), and a hotter
region being traced by \ion{O}{vi} (peaking at $10^{5.5}\,\text{K}$).
On the other hand, the detected low ionization species of \ion{O}{i},
\ion{C}{ii} and \ion{Si}{ii} trace WNM and WIM gas with a temperature
$<10^{4}\,\text{K}$ as does \ion{H}{i}.

Table~\ref{tab:LowIonLines} summarizes the parameters of our observation-based
\ion{H}{i} model components (calculated along the same sight line)
and measured components of three of the low ions from FSW10. The core
component central velocities line up quite well within a few $\text{km\,\ s}^{-1}$
of the \ion{H}{i} and the wider component central velocities are
also quite close, but have a wider range. The velocity dispersions,
$\sigma_{v}$ of the core components are all comparable, but those
of the envelope velocity are widely scattered. We interpret these
data in the following sections. 
\begin{deluxetable*}{lcccccc}  % <--- column justification (center/left/right)
\tablecolumns{7}
\tablewidth{0pt}
\tabletypesize{\small}
\tablecaption{Low Ion Line Properties toward NGC~7469\label{tab:LowIonLines}}
\tablehead{   % column headings
\colhead{} & \colhead{} & \multicolumn{2}{c}{Core} & \colhead{} &
\multicolumn{2}{c}{Envelope} \\ 
  \cline{3-4} \cline{6-7}\\
  \colhead{} &
  \colhead{$E_{i+1}$\tablenotemark{a}} &
  \colhead{$v_0$} &
  \colhead{$\sigma_v$} &
  \colhead{} &
  \colhead{$v_0$} &
  \colhead{$\sigma_v$}\\
  \colhead{Line} &
  \colhead{(eV)} &
  \colhead{(km~s$^{-1}$)} &
  \colhead{(km~s$^{-1}$)} &
  \colhead{} &
  \colhead{(km~s$^{-1}$)} &
  \colhead{(km~s$^{-1}$)}
}
\startdata
{\ion{H}{i} 21 cm\tablenotemark{b}} & {13.6} & {$-336\pm0.2$} & {$9.7\pm5.0$} & {} & {$-357\pm0.4$} & {$28\pm5$}\\
{\ion{O}{i} $\lambda1302.168$\tablenotemark{c}} & {13.6} & {$-337\pm1$} & {$5.6\pm1.4$} & {} & {$-364\pm1$} & {$4.2\pm4.2$}\\
{\ion{Si}{ii} $\lambda1260.422$\tablenotemark{c}} & {16.3} & {$-335\pm1$} & {$7.1\pm1.4$} & {} & {$-368\pm2$} & {$9.2\pm2.8$}\\
{\ion{C}{ii} $\lambda1334.532$\tablenotemark{c}} & {24.4} & {$-330\pm1$} & {$9.2\pm1.4$} & {} & {$-360\pm1$} & {$16.3\pm2.8$}\\
\enddata
\tablenotetext{a}{Ionization potential to next level.}
\tablenotetext{b}{This work, mean model values $2.5\,'$ off-axis (toward NGC~7469) with $9.1\,'$ beam.}
\tablenotetext{c}{FSW10 Table~1, $b$ values scaled to $\sigma_v$, aperture is $0.2\times0.06\,''$.}
\end{deluxetable*}

\subsection{Cloud Envelope}

\label{sub:Discussion:Envelope}

\subsubsection{Smooth conduction-dominated boundary layer}

The detected envelope \ion{H}{i} line has an average modeled velocity
dispersion $\sigma_{ve}=29\,\text{km\,\ s}^{-1}$ (Table~\ref{tab:ModCloudProp})
which does not vary significantly across its full extent, which is
$2.6\, r_{c}$ (Figure~\ref{fig:CloudModelParms}). Although the
observed velocity FWHM might suggest a slight increase away from the
cloud center, this is not well constrained with our angular resolution.
This dispersion is $\sim3$ times higher than what is expected for
the WNM/WIM gas at $\sim10^{4}\,\text{K}$ and could be interpreted
as being due to gas at kinetic temperature $T<1.0\times10^{5}\,\text{K}$
with an additional turbulent component. One possible explanation is
that the envelope represents a conduction-dominated (evaporative)
boundary layer, along the lines of \citet{cowie77}, in relatively
smooth transition from the $\sim10^{4}\text{K}$ WNM/WIM core to the
$\sim10^{6}\text{K}$ WIM Halo. We do not consider radiative effects
on the interface as analyzed by \citet{mckee77} since the estimated
core radius of 147~pc (Table~\ref{tab:ModCloudProp}) is well below
the critical radiation radius, $R_{rad}$ defined in that analysis
as the radius above which radiation effects become significant. The
analysis, based on cooling at solar metallicity yields $R_{rad}>1.6\,\text{kpc}$
for an assumed final (Halo) temperature $T>10^{6}\,\text{K}$ and
density $n<10^{-4}\,\text{cm}^{-2}$ \citep{sembach03}. This is already
much larger than our estimated cloud radius. The lower metallicity
in this part of the MS of $\sim0.1\,\text{solar}$ \citep{fox10}
implies a lower cooling rate. Thus the mean free path in the boundary
layer, represented by the scale length $\lambda(T)$, increases \citep[Equation 7]{mckee77}
making for an even higher critical radius. This makes the non-radiative
assumption even more comfortable.

Interpreting the envelope as single-phase gas in a conduction-dominated
boundary layer, we first estimate the maximum detectable gas temperature
that would leave sufficient neutral fraction to produce the observed
$2.1\times10^{18}\,\text{cm}^{-2}$ neutral envelope column density
(Table~\ref{tab:ModCloudProp}) through a realistic column depth.
A rough equilibrium calculation was performed for Hydrogen using an
approximate collisional ionization formula \citep[Equation 13.11]{draine11}
and interpolated recombination rate coefficients \citep[Table 5.2]{spitzer98}
under optically thin conditions. It showed that envelope gas of density
$<0.011\,\text{cm}^{-3}$ (twice the core neutral density of Table~\ref{tab:ModCloudProp})
at a temperature $>3.8\times10^{4}\,\text{K}$, would be so highly
ionized as to require a column $>140\,\text{kpc}$ (twice the assumed
distance) to obtain the measured column density. So, the observed
envelope would have to trace gas no warmer than $3.8\times10^{4}\,\text{K}$
, with kinetic (thermal) dispersion $\sigma_{k}<18\,\text{km\,\ s}^{-1}$
. Given the kinetic component, the turbulent component required to
produce the total dispersion would have $\sigma_{t}>23\,\text{km\,\ s}^{-1}$.

Although no conduction-dominated analyses or simulations that we are
aware of consider intrinsically turbulent cloud structure, they predict
the temperature profiles of static (evaporative/condensing) boundary
layers. \citet{dalton93} considered both classical and saturated
heat flux and derived temperature profiles as a function of $r/r_{c}$
and the global saturation parameter, $\sigma_{0}$. In \citet{gnat10}
Figure~1, these profiles are plotted for a range of $\sigma_{0}$
values. In our case, we assume $T_{HIM}=T_{Halo}>10^{6}\,\text{K}$
and for our estimated maximum envelope temperature, we have $T/T_{HIM}<3.8\times10^{4}/10^{6}=0.038$ where
$r/r_{c}$ gets only slightly larger than one, regardless of saturation
parameter. This would make the transition to our maximum envelope
temperature on order of $0.01\, r_{c}$. In fact, even if the above
detectability analysis is ignored and assuming the envelope dispersion
is completely thermal, with corresponding $T_{max}=1.0\times10^{5}\,\text{K}$,
the transition distance is still only on order of $0.1\, r_{c}$.
This is clearly inconsistent with our extensive detected envelope
which reaches a distance of $2.6\, r_{c}$ beyond the core. 

\citet{vieser07b} performed simulations that included classical and
saturated heat flux (their model R3) for a cloud of similar total
mass to ours and under ambient temperature and density on the same
order as the Halo. Although nearly sonic flows and a thin turbulent
layer ($\sim5\%$ of $r_{c}$) developed in these simulations, no
systemic turbulence developed nor was it considered. The initial cloud
radius, at $41\,\text{pc}$, is less than $1/3$ the size of our cloud,
but we assume that things roughly scale with radius and ambient temperature
as do the derived profiles of \citet{dalton93}. A temperature profile
is not provided, but if we interpret the density profile of their
Figure~12 assuming roughly constant pressure, the steep drop in density
from $r/r_{c}\approx1$ to about 10 times that of the ambient Halo
at $r/r_{c}\approx1.2$ suggests the temperature at that point would
be $T\sim10^{5}\,\text{K}$ for $T_{Halo}\sim10^{6}\,\text{K}$. From
this temperature profile, our detected envelope gas should extend
no further than $20\%$ of core radius (again assuming that the velocity
dispersion is purely thermal). This is much further than the \citet{dalton93}
profiles predict, but still less than an order of magnitude as deep
as our observed envelope, which extends as far as $260\%$ of core
radius, even after convolution with the GBT beam is taken into account.
\citet{vieser07b} also simulated the cloud with self-gravity (their
model R4) and then again with cooling/heating added (their model R5),
with each compressing the transition range further. We can therefore
state that our observed envelope extends further, \emph{by more than
an order of magnitude}, than simulated or derived results predict
for an evaporating cloud without systemic turbulence. The impact that
turbulence throughout our core and envelope would have on these predictions
is not known, yet one would expect it to be significant. However,
an order of magnitude increase in the extent of the low temperature
range of the boundary layer seems unlikely.

\subsubsection{Turbulent mixing layer}

As an alternative interpretation, the large velocity dispersion of
the cloud envelope may be the result of the superposition of multiple,
more or less independent clumps of WNM/WIM gas turbulently intermixed
in the boundary layer with warmer phase gas along the lines of the
Turbulent Mixing Layer (TML) described in \citet{begelman90}. This
approach provides a simpler and more supportable explanation for our
observations. In this picture the broad velocity width of the envelope
is due to the superposition of discrete clumps of $\sim10^{4}\,\text{K}$
WNM gas (as found in the cloud core, Table~\ref{tab:ModCloudProp}),
photoionized by the background and Galactic UV radiation, and turbulently
dispersed in a mixture with much warmer gas approaching HIM temperatures
with $T\sim10^{5}\,\text{K}$. The superposition of multiple WNM ($\sim10^{4}\,\text{K}$)
clumps, each with $\sigma_{vc}\sim10\,\text{km\,\ s}^{-1}$ turbulently
dispersed with $\sigma_{t}\sim27\,\text{km\,\ s}^{-1}$ will produce
an approximate Gaussian shape with $\sigma_{ve}\sim29\,\text{km\,\ s}^{-1}$,
similar to the observed total envelope dispersion.

The low metal ion envelope velocity dispersions of Table~\ref{tab:LowIonLines}
should also be consistent with this picture. \ion{O}{i} and \ion{Si}{ii}
envelope components have comparable dispersion, but are very low compared
to \ion{H}{i}. \ion{C}{ii} has significantly higher dispersion
than \ion{O}{i} and \ion{Si}{ii}, yet less than \ion{H}{i}.
It is important to emphasize that \ion{H}{i} is observed with a
$9.1\,'$ beam and averaged all around the cloud's vicinity through
the spatial averaging we applied, so it would sample a large number
of these WNM clumps over a wide range of dispersed velocities. On
the other hand, the metal ions are observed with a ``pencil'' beam
through the center of the cloud, passing through a much smaller number
of the clumps and having a smaller dispersion. The larger \ion{C}{ii}
dispersion could be explained by its significantly higher ionization
potential allowing it to survive into warmer and more turbulent gas
surrounding the WNM clumps.

The TML model, first suggested by \citet{begelman90}, explains turbulent
mixing of cool and hot gas arising from the KH or shear instability
induced by the velocity difference across the interface between the
two types of gas. The TML consists of a continuum of phases between
the cooler $\sim10^{4}\,\text{K}$ (WNM/WIM) stripped gas and intermediate
temperature gas at $\bar{T}=\sqrt{T_{cool}T_{hot}}\sim10^{5}\,\text{K}$
(assuming $T_{hot}=T_{Halo}\sim10^{6}$ K) coexisting in a turbulent
mixture. Also, since cooling is very efficient at this intermediate
temperature, they suggest that some of the ablated gas would rejoin
the cooler phase rather than be lost to the hot phase gas, a possible
means for slowing the ablation process and extending cloud lifetime.

Hydrodynamic simulations of WNM gas clouds passing at supersonic velocities
through the HIM provide further qualitative support to our observations.
In simulations, WNM clouds consistently develop a hierarchy of smaller
clumps peeling off the cloud as a result of KH instability, similar
to the picture of WNM/WIM clumps we propose \citep{esquivel06,vieser07a,heitsch09,kwak11}.
In the present case, our cloud core is probably a very large clump
recently separated from the western complex of Figure~\ref{fig:ClumpMaps}
as discussed above. Another consistent feature of the simulations
is the development of vorticular flow along the shearing interface
of the main cloud as well as clumps and protuberances that develop
there. A most striking example are the early evolution simulations
shown in Figure~2 of \citet{vieser07a} where vortices are fully
developed in the non-conductive case and appear to be forming in the
more slowly developing conductive case. In the longer-term multi-ion
simulations of \citet{kwak11} the material is tracked as the developing
TML entrains it. The simulation represented in their Figure~2 extend
to much larger times and the cloud has properties more like our cloud
than the simulations of \citet{esquivel06} or \citet{vieser07a}.
As the gas falls behind the HVC it mixes with the ambient gas producing
intermediate temperature gas. The temperature and velocity structure
of the ablated material increases in complexity and range with time.
At later times, the simulations indicate a wide $10^{3}$ to $10^{5.5}\text{K}$
temperature range in the ablated material mixture with $\sim40\,\text{km\,\ s}^{-1}$
of velocity range in the $\sim10^{4}\,\text{K}$ gas near the cloud
axis. This is similar to the observed velocity dispersion of the \ion{H}{i}
envelope. Also significant, although not easy to discern in Figure~2
of \citet{kwak11}, is the vorticular transport of gas from the mixing
layer to the central region where it reverses direction with respect
to the entrained gas as discussed by the authors. They also describe
it as a means by which gas cooled in the mixing layer replenishes
the cooler gas at the center as suggested by \citet{begelman90} and
mentioned above. This vorticular motion in the simulations is consistent
with our proposed picture of clumps dispersed in both leading and
lagging velocities with respect to the cloud in our observed envelope.

\subsection{Cloud Core}

\label{sub:Discussion:Core}

The observed \ion{H}{i} core component, with its mean line width
consistent with $T<1.2\times10^{4}\,\text{K}$, is reasonably interpreted
as the neutral component of a central concentration of moderately
turbulent WNM to WIM supplying cooler gas to the TML after being stripped
from its periphery. The kinetic temperature $T$ and turbulent dispersion
$\sigma_{t}$ of the gas can be estimated from the line velocity dispersions,
$\sigma_{1}$ and $\sigma_{2}$ of two co-spatial species with atomic
masses $m_{1}$ and $m_{2}$ by solving the two equations:
\begin{align}
\frac{k_{b}T}{m_{1}}+\sigma_{t}^{2} & =\sigma_{1}^{2}\label{eq:temp-turb-1}\\
\frac{k_{b}T}{m_{2}}+\sigma_{t}^{2} & =\sigma_{2}^{2}.\label{eq:temp-turb-2}
\end{align}
 More than one species could be incorporated by finding an optimum
solution for a set of equations for all co-spatial species, of course.

The components of the three metal species at or near the \ion{H}{i}
core velocity listed in Table~\ref{tab:LowIonLines} should sample
a line of sight through the WNM core and the (photoionized) WIM envelope
that probably surrounds it. However, they may not all be strictly
co-spatial. The density profile of each species can be quite different
along the line of sight due to different ionization potentials responding
to a decreasing UV flux toward the shielded core center. \ion{H}{i}
and \ion{O}{i} have the lowest ionizing potentials and so survive
mostly toward the center and concentrate there. \ion{Si}{ii} and
\ion{C}{ii} have much higher potentials and, surviving the higher
flux in the periphery, are distributed more evenly throughout the
core. Unless gas along the line of sight has a uniform temperature
and turbulent dispersion out to a distance where \ion{Si}{ii} and
\ion{C}{ii} are mostly depleted, these two ions would be biased
with respect to \ion{H}{i} by sampling gas near the likely warmer
and more turbulent periphery near or even into the TML. \ion{H}{i}
and \ion{O}{i} have nearly identical ionization potential which
leads to nearly identical ion fraction profiles, sampling points similarly
along the line of sight, heavily weighted toward the self-shielded
center of the cloud. As a result, we can use \ion{H}{i} and \ion{O}{i}
to separate the thermal and turbulent components and be confident
that the result reflects the mean along the line of sight, weighted
heavily toward the center of the core and not affected significantly
by the TML.

Using the velocity dispersion values for \ion{H}{i} and \ion{O}{i}
from Table~\ref{tab:LowIonLines} and solving the above equations
yields $T=8350\pm350\,\text{K}$ and turbulent dispersion $\sigma_{t}=5.3\,\text{km\,\ s}^{-1}$,
both of which are quite consistent with expectations for WNM/WIM.
It is interesting to note that if we assume that \ion{Si}{ii} and
\ion{C}{ii} are at this same temperature throughout the core column,
their turbulent dispersion components come out to $6.9\,\text{km\,\ s}^{-1}$
and $8.9\,\text{km\,\ s}^{-1}$, respectively. This is consistent
with \ion{Si}{ii}, with its ionization potential higher than \ion{H}{i},
surviving further into an increasingly turbulent periphery and \ion{C}{ii},
with an even higher potential, surviving further still into even higher
turbulence as the WNM/WIM core transitions to the TML.

\subsection{Cloud Ionization}

\label{sub:Discussion:Ionization}

Our picture of the cloud is of a partially ionized stable WNM/WIM
core, surrounded by a TML, in which diffuse WNM/WIM clumps are dispersed.
These small, dispersed clumps are exposed to not only ionizing extragalactic
and Galactic radiation, but are also embedded in the ionizing radiation
produced locally by the collisional processes in the warmer phases
of the TML. We therefore assume that the clumps are highly ionized
and essentially transparent to Galactic UV and extragalactic background,
which reaches the core at full strength. We now investigate whether
a 3D photoionization model of the cloud core exposed to the Galactic
and extragalactic UV radiation field can reproduce a neutral component
consistent with our core model. We use a version of the 3D Hydrogen-only
ionization equilibrium code of \citet{wood00}. The code was set up
to include a Galactic and extragalactic ionizing flux. A 3D spherical
cloud of uniform volume density is discretized onto a linear Cartesian
grid with 129 cells on a side. The gas is assumed to be isothermal,
with recombination rates based on a kinetic temperature of 8000~K.

The gas can be made clumpy with a fractal-generated density modulation
calculated using the algorithm of \citet{elmegreen97} (as described
in several papers \emph{e.g.}, \citealt{mathis02,wood05}). This algorithm
leaves a fraction of the gas as the smooth component and redistributes
the remaining gas into hierarchical clumps. The fractal dimension
$f$ and parameters $N_{1}$ through $N_{5}$ govern the process.
For more on these parameters and their effect on density structure,
see the above-mentioned references. In this study, our purpose is
only to include the general effect of clumpiness on the density profile
shape for no other reason than to add some measure of realism to the
cloud. It is not to emulate large-scale structure (the original purpose
of this algorithm) nor any particular ISM environment. We empirically
adjusted the $f_{smooth}$ and fractal parameters of the algorithm
to produce, after ionization, only one distinct neutral structure
representing the cloud core (with some parameter settings, there were
two cores) while still producing distinct substructure. We settled
on $f_{smooth}=0.85$, fractal dimension $f=2.6$ with $N_{1}$ through
$N_{5}$ set to 64, 16, 16, 16 and 16, respectively.

\begin{figure*}
\begin{centering}
\subfloat[]{\begin{centering}
\includegraphics[width=8cm]{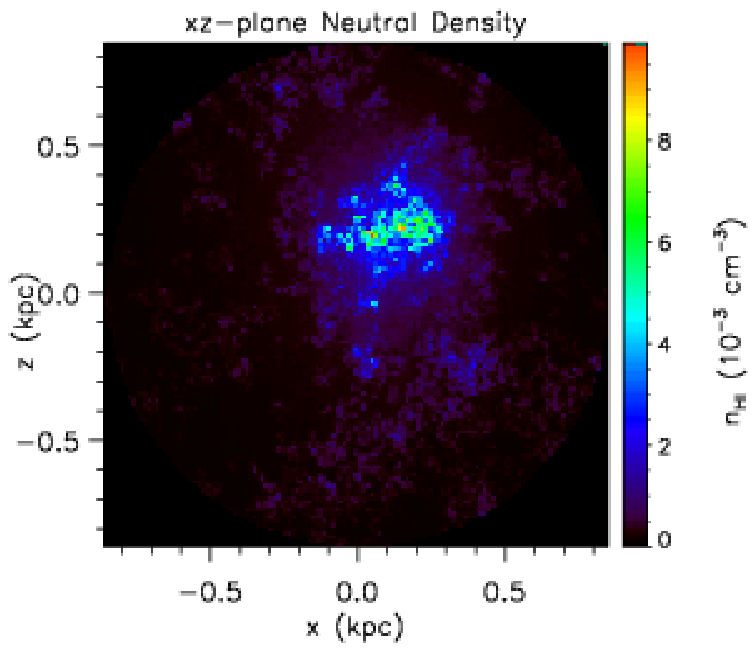}
\par\end{centering}

}
\par\end{centering}

\begin{centering}
\subfloat[]{\begin{centering}
\includegraphics[width=7.5cm]{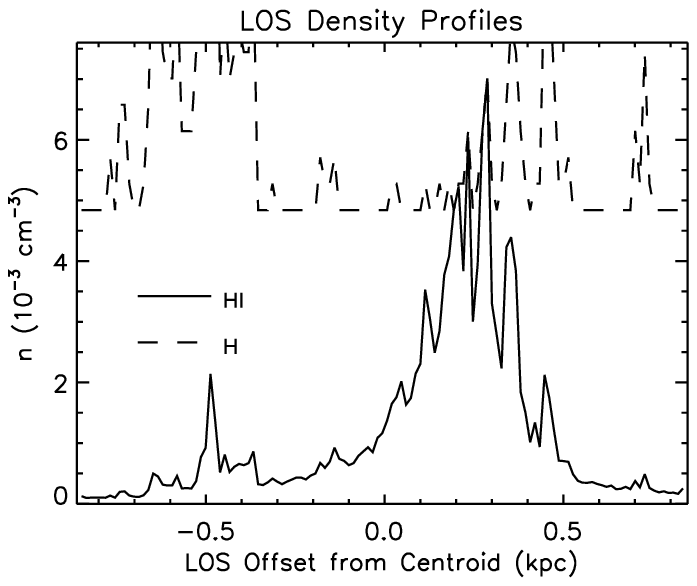}
\par\end{centering}

}\subfloat[]{\begin{centering}
\includegraphics[width=7.5cm]{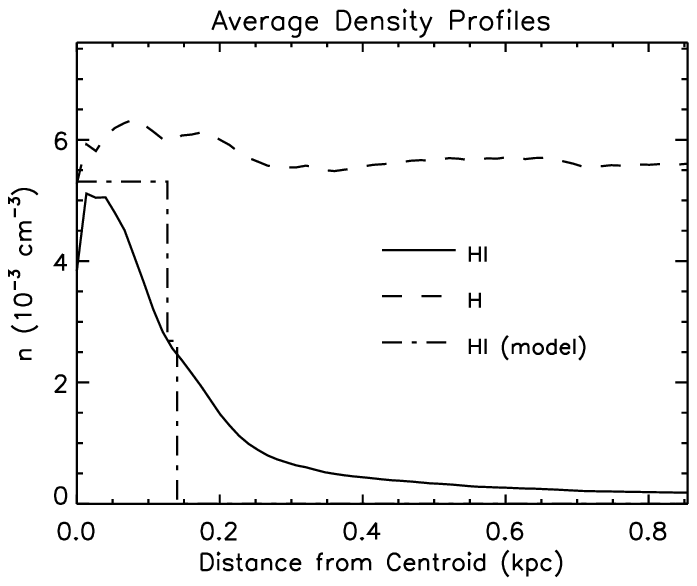}
\par\end{centering}

} 
\par\end{centering}

\caption[Ionization modeling results.]{Neutral density characteristics of a 15\% clumpy cloud core with
uniform average density $5.5\times10^{-3}\,\text{cm}^{-3}$, ionized
3D spherical cloud at Galactocentric distance $70\,\text{kpc}$ exposed
to standard Galactic and extragalactic ionizing radiation assuming
the surrounding envelope is highly ionized and transparent. The cloud
radius was adjusted to achieve the central column density of the cloud
core observation-based model, $N_{HI}=4.8\times10^{18}\,\text{cm}^{-2}$.
Plots for each case are (a) LOS neutral column density; (b) density
profile along a LOS through the centroid; and (c) average radial density
profile along with that from our model. \label{fig:IonizedCore}}
\end{figure*}

In order to demonstrate that photoionization alone can produce a neutral
core consistent with our observations, we exposed a spherical cloud
of uniform average density to ionizing radiation appropriate to our
assumed distance of $70\,\text{kpc}$. A density of $5.5\times10^{-3}\,\text{cm}^{-3}$
was assumed which is somewhat higher than that of our core model (Table~\ref{tab:ModCloudProp}), thus
allowing for a moderate ion fraction at the center. Again this is
merely a demonstration illustrating what the actual profile, unresolved
by our observations, might look like based on exposure to a realistic
ionizing field. The ionization flux used is a combination of constant
isotropic extragalactic ionizing background of $7.9\times10^{3}\,\text{cm}^{-2}\,\text{s}^{-1}$
from \citet{faucher09}%
\footnote{Calculated from energy spectrum provided online by the authors at
www.cfa.harvard.edu/\textasciitilde{}cgiguere/UVB.html%
} and directional Galactic flux of $1.3\times10^{4}\,\text{cm}^{-2}\,\text{s}^{-1}$
from the model of \citet{fox05}%
\footnote{From data kindly provided by Joss Bland-Hawthorn.%
} calculated at the cloud's Galactic coordinates from Table \ref{tab:ModCloudProp}
and assumed LOS distance of $70\,\text{kpc}$. The results are shown
in Figure~\ref{fig:IonizedCore}. To obtain the ionization results,
we kept the density and fractal parameters constant and adjusted the
cloud radius in order to obtain the observed central \ion{H}{i}
column density of $N_{c}=4.8\times10^{18}\,\text{cm}^{-2}$ (Table~\ref{tab:ModCloudProp})
as measured along the Galactocentric LOS, defined in the local rectilinear
coordinate system as the $z$-direction. When calculating average
density profile and column density, the cloud center is determined
from the 3D centroid of the neutral density distribution. 

Figure~\ref{fig:IonizedCore}(a) shows the resultant neutral density
structure in the $x-z$ plane after the cloud has reached ionization
equilibrium. Figure~\ref{fig:IonizedCore}(b) shows the cloud \ion{H}{i}
and total Hydrogen density profile through the centroid along the
LOS. The photoionized cloud component is more extended than the neutral
core. Note the asymmetry with the neutral core to right of center.
This is due to the directional Galactic radiation component, providing
increased flux along the LOS, which is aligned closely with Galactic
Center. The remaining neutral core has a FWHM size of 319 pc. Figure~\ref{fig:IonizedCore}(c)
shows the average radial \ion{H}{i} and total Hydrogen density
profiles from the simulation. Also shown is our observation-based
model core \ion{H}{i} profile for comparison (from Table~\ref{tab:ModCloudProp}).

The neutral density profile (Figure~\ref{fig:IonizedCore}(b)) of
the ionized cloud core approximates a composite exponential profile
dropping rapidly near the more neutral center and then a longer scale
takes over as it becomes more ionized at larger radii. This is typical
of ionization fronts, as in \citet{zheng02}, where logarithmic plots
show three slope regimes, a flat range if the core is fully shielded,
followed by a steep region with moderate ionization and then a shallower
one with high ionization. Since in our example the core is not fully
shielded, there is no flat region. Our crude core model of constant
density derived from observations approximates this more realistic
density profile as a constant, roughly equal to the central density
truncated at about half the full extent of the core, which has a radius
of $319\,\text{pc}$.

This phoionization model also provides information about the neutral
fraction of the cloud. We calculate the neutral fraction from Figure~\ref{fig:IonizedCore}(b)
by comparing neutral and ionized hydrogen profiles and find a fraction
of $24\%$ along the central LOS. This is three times higher than
$7\%$ determined by FSW10 based on CLOUDY simulations for low ions.
FSW10 also determined a lower limit for the incident ionization parameter,
$U=n_{\gamma}/n_{H}>5\times10^{-4}$. The equivalent $U$ for our
example is $9.7\times10^{-5}$, a factor of 5 lower. To get the FSW10
value of $U$, our simulations must be at an either unrealistically
low or high distance ($<20\,\text{kpc}$ or $>200\,\text{kpc}$) to
get high enough Galactic $n_{\gamma}$ or low enough $n_{H}$, respectively.
Realistically, the ionizing flux and therefore $U$ would be even
lower if partial shielding of the extragalactic flux by the Galactic
disk and the neighboring western complex were taken into account,
increasing the discrepancy. Given the insight we now have into the
structure along the NGC~7469 sight line, it can be argued that for
the core and envelope structure we propose, FSW10's CLOUDY analysis
overestimates both $U$ and the ion fraction for the low ion core
environment. Their analysis used the total metal column densities,
which include significant metal ion column density from envelope velocity
components, as targets for the simulation. If these absorption components
occur in highly ionized WIM fragments dispersed in the TML as we propose,
their contributions would not be well represented as part of an integral
CLOUDY slab model. Also, higher potential ions like \ion{Si}{ii}
and \ion{C}{ii} would disproportionately sample warmer portions
of the ambient collisional mixture. Furthermore, local TML collisional
FUV radiation could contribute to the ionization state of the WIM
fragments, further biasing the modeling results. It would be interesting
to repeat the FSW10 simulations using only the metal column densities
associated with the core velocity as targets. Given the high slopes
in FSW10 Figure~5 (left) for the high potential ions, this could
result in a significantly lower $U$ and ion fraction more in line
with our results.

In summary, we showed that we can reproduce the observed \ion{H}{i}
column density and the rough size of the cloud core relatively easily
and with simple assumptions, in the scenario where the core represents
the remnant of a larger cloud exposed to Galactic and extragalactic
background ionizing radiation. The phoionization exercise also suggests
that a large percentage ($76\%$ in this example) of the warm phase
core of this cloud is ionized.

\subsection{Cloud Lifetime}

\label{sub:Discussion:Lifetime}

If we assume that the mixing layer is merely a transport mechanism
for the WNM/WIM of the cloud to ablate into the Halo, we can apply
the hydrodynamic ablation equations of \citet{pittard07} (pg. 245)
to estimate the lifetime of the model cloud. In this calculation,
we use the total estimated cloud mass from Table~\ref{tab:ModCloudProp}
which includes all warm phase gas (WNM or WIM) and the lifetime is
the time required for all of this warm phase gas to be heated and
absorbed into the hot Halo. This lifetime estimate will be on the
high side since we are starting with the ablation process well underway
with the warm envelope mass already in transition. It should be noted
that our averaging techniques expose a normally undetected envelope
whose observed neutral mass is $\approx7$ times that of the core,
so a very large portion of the mass of this cloud is in the envelope.
This suggests that ablation time calculations based on cloud masses
inferred from column densities of conventional \ion{H}{i} maps will
be quite low since the unobserved envelope may contain the majority
of unablated mass.

Assuming a distance $D=70\,\text{kpc}$, the total cloud mass estimate
from Table~\ref{tab:ModCloudProp} is $M_{c}=4\times10^{5}M_{\odot}$.
We assume a peculiar cloud velocity of $400\,\text{km\,\ s}^{-1}$
with respect to a non-rotating Halo ($v_{LOS}=350\,\text{km\,\ s}^{-1}$
plus some moderate proper motion), WNM/WIM cloud sound speed $c_{c}\approx10\,\text{km}\,\text{s}^{-1}$,
and Halo properties $T_{halo}\approx10^{6}\,\text{K}$, and $n_{halo}\approx10^{-4}\,\text{cm}^{-3}$
\citep{fang06} on order of typically assumed values. The flow will
be supersonic for these conditions, so we use $\dot{M}=(M_{c}c_{c})^{\frac{2}{3}}(\rho_{halo}v)^{\frac{1}{3}}\approx3\times10^{3}\, M_{\odot}\,\text{Myr}^{-1}$
for a cloud lifetime of $t_{c}\approx140\,\text{Myr}$. The infall
distance prior to destruction is on order of $v_{LOS}t_{c}=50\,\text{kpc}$,
not quite surviving to the Galactic Disk from the assumed distance.
However, mass loss rates could be considerably lower due to factors
not accounted for in the above equation such as recycling of WNM/WIM
gas through rapid cooling \citep{begelman90,kwak11}, conduction effects
\citep{vieser07a} and magnetic fields \citep{esquivel06}.

\citet{kwak11} performed long term simulations of clouds with properties
on the same order as ours except for an assumption of solar metallicity
rather than the 0.1~solar estimate for this cloud \citep{fox10}.
Their model ``D'' has comparable initial mass and retains 70\% of
its mass after $240\,\text{Myr}$. Its velocity is less than half
that of ours, but the simulations and the authors' analysis suggest
that once supersonic, flow velocity has a large effect on cloud morphology,
but little effect on the mass loss rate. These models suggest the
cloud should have a lifetime as WNM/WIM on order of twice that calculated
with the \citet{pittard07} equation, long enough for it to potentially
get within reach of the Galactic Disk. It should be pointed out that
if cooling in the TML indeed slows ablation through recycling as mentioned
above, the lower metallicity would reduce the lifetime by some unknown
amount.

On the other hand, \citet{heitsch09} performed 3D simulations of
clouds falling through the Halo assuming 0.1~solar abundances so
cooling effects are appropriate to our cloud. They directly traced
the smaller WNM/WIM clump fragments like those that compose our observed
envelope. The cases illustrated are mostly subsonic at all times and
clumps generally trail the core mass, but in one of the free-fall
simulations shown, it briefly becomes transonic and develops some
leading clumps suggesting that vortices are developing as they do
in other supersonic simulations like \citet{kwak11}. Their model
``Wb1a15b'' has initial mass $3\times10^{3}\,\text{M}_{\odot}$
and loses $50\%$ of its mass to the Halo in $50\,\text{Myr}$ (their
Figure~3) and extrapolating to full mass loss, lasting $\sim100\,\text{Myr}$.
To estimate our cloud's lifetime, we start with the integral core
mass only (without the envelope) as this corresponds to the initial
state of the simulations. We assume that the core will ablate similarly
to the simulated cloud independent of a the pre-existing envelope.
Our neutral core mass at $70\,\text{kpc}$ is $2.5\times10^{3}\,\text{M}_{\odot}$
(Table~\ref{tab:ModCloudProp}) and allowing for $24\%$ ionization
as estimated in Section~\ref{sub:Discussion:Ionization}, is $\sim3\times$
the mass of the simulated cloud. The flow is $\text{Mach}\,0.7$ in
the simulation and so should ablate at nearly the same rate. The extrapolated
lifetime of our cloud would be on order of $300\,\text{Myr}$, more
than twice the \citet{pittard07} calculation, and easily sufficient
to reach the Disk with some WNM/WIM mass remaining.

In summary, the low-end estimate of cloud survival time as WNM/WIM
prior to being subsumed by the Halo is $140\,\text{Myr}$, insufficient
to reach the Galactic Disk. This is based on the analysis of \citet{pittard07}
which does not account for cooling or other factors that have been
shown to suppress ablation. The numerical results of \citet{kwak11}
which include cooling, suggest that the time to full destruction of
warm gas in a similar cloud is much longer, by up to a factor of 2
or more although assumed solar metallicity puts this estimate on the
high side. However, rough extrapolation of \citet{heitsch09} simulations,
with abundances appropriate to our cloud indicate that, at more than
twice the \citet{pittard07} lifetime, it is possible or even likely
that warm gas will reach the Disk.

\section{Summary and Conclusions}

\label{sec:Conclusions}

We have obtained deep \ion{H}{i} 21~cm emission measurements mapping
the vicinity of a relatively isolated, circularly projected cloud
in the northern MS and applied a new spatial averaging technique to
profile its mean spectrum vs. projected distance from the cloud center.
Although near a larger cloud complex from which it appears to be separating,
the cloud is well defined in angular extent and very well defined
in velocity. By exploiting the approximate azimuthal symmetry of the
cloud and averaging along successively larger concentric circular
paths, we obtained correspondingly increased sensitivity without compromising
the essential spatial resolution along the radial dimension. With
this method, we detected and characterized core emission having $\approx20\,\text{km\,\ s}^{-1}$
WNM/WIM-like FWHM with velocity of $\approx-340\,\text{km\,\ s}^{-1}$
and an envelope with broad diffuse emission having $\approx60\,\text{km\,\ s}^{-1}$
FWHM and velocity of $\approx-360\,\text{km\,\ s}^{-1}$ extending
well beyond the apparent periphery of the cloud's emission. We then
obtained robust estimates of the core and envelope spectral properties
by optimally matching the observed, averaged profile to a simple,
spherical 3D core-plus-envelope emission model of the cloud projected
on to the sky and averaged using the same method.

The envelope probes the neutral component of a boundary layer between
the infalling WNM/WIM cloud core and the ambient HIM of the Halo.
Assuming that the broad line width reflects very warm transition gas
in a conductive boundary layer, we estimate the detected extent of
the neutral envelope is more than an order of magnitude beyond what
should be detectable based on the analyses and simulations of cloud
evaporation. We conclude that a Turbulent Mixing Layer (TML) best
explains the observed characteristics. Theoretical treatments and
numerical simulations in the literature show that a TML has properties
that can explain both the envelope line width and leading velocity
where the envelope is composed of small clumps of narrower-width WNM/WIM
dispersed over a large velocity range amidst warmer gas in the TML.
This presents a wide Gaussian-like spectrum when a large number or
these small clumps are within the beam with some leading and some
lagging the core component. The leading clumps are also explained
by the TML where vortices develop near the infalling core when flow
is supersonic as expected and demonstrated in various numerical simulations.

Passing through the cloud and fortuitously near its center, is the
NGC~7469 background source sight line which was characterized by
FSW10 using optical and UV absorption data. Because \ion{O}{i}
and \ion{H}{i} trace virtually identical profiles weighted toward
the center of a partially photoionized cloud, we used measured velocity
dispersions from FSW10's \ion{O}{i} and our \ion{H}{i} core
velocity components to establish a core temperature $T=8350\pm350\,\text{K}$
with a turbulent component $\sigma_{t}=5.3\,\text{km\,\ s}^{-1}$,
consistent with WNM/WIM.

The role of photoionization on the cloud was investigated using a
3D photoionization equilibrium program. Assuming that the collisionally
ionized warm TML gas as well as the interspersed WNM/WIM clumps are
highly ionized, we treated the envelope as transparent to Galactic
and extragalactic ionizing radiation. We exposed a clumpy spherical
cloud with radius $320\,\text{pc}$ at a distance of $70\,\text{kpc}$
with uniform average Hydrogen density of $5.5\times10^{-3}\,\text{cm}^{-3}$
to combined Galactic and extragalactic ionizing flux based on published
values at this distance. We determined that the centrally concentrated
exponential profile of the neutral component is approximated well
by our simple uniform neutral density model of $\sim1/2$ the radius.
The core's actual extent as WNM/WIM is therefore on order of twice
that of our 3D neutral component model. The neutral fraction along
the central LOS is $24\%$ and the ionization parameter is $9.7\times10^{-5}$,
quite different from FSW10's estimates of $7\%$ and $>5\times10^{-4}$,
respectively. We are unable to approach FSW10's values with any realistic
cloud at a realistic distance. We conjecture that the discrepancy
is at least partially due to FSW10's inclusion of what we suggest
are envelope velocity components in abundance calculations. If produced
in the clumpy, warmer and partially collisional envelope environment
we propose, these components would not be represented well by CLOUDY.

The cloud lifetime assuming simple hydrodynamic ablation driven by
shear instability in supersonic flow is estimated at $\sim140\,\text{Myr}$,
insufficient for any WNM or WIM to reach the Disk and potentially
fuel future star formation. However, a TML is expected to partially
suppress ablation through re-cooling of some of the heated gas \citep{begelman90}.
Simulation of a comparable infalling cloud by \citet{kwak11} which
includes cooling indicates a dramatically increased $\sim2\times$
lifetime, suggesting that a large portion of the total mass could
reach the disk. This result is supported by extrapolating the simulations
of \citet{heitsch09}.

Through our use of the spatial averaging technique, this work provides
the first look (or more accurately, a peek) at the structure of the
highly ionized and diffuse boundary layer between the warm neutral
gas of the MS and the HIM of the Halo from the perspective of the
low temperature side of the interface. FSW10 provides a rich set of
data and analysis for the fortuitously aligned NGC~7469 sight line,
creating a mini-laboratory to study MS-Halo interaction. We were able
to incorporate some of their results to complement ours, but were
unable to quantify characteristics of the barely probed TML. To fully
exploit this alignment of complementary data, hydrodynamic simulations
that emulate the evolution of the larger environment of the cloud
(\emph{i.e.}, including the neighboring cloud complex) would be needed
that allow simulated observations on a similar cloud feature. Photoionization
would have to be accounted for and its possible impact on the hydrodynamic
evolution would have to be addressed.

Future work will include applying the technique to isolated circular
or filamentary structures (with or without benefit of well-placed
background sources) in other parts of the MS, both upstream and downstream
to see if this apparent mixing layer is common and if its characteristics
vary with time, indicated by the location along the MS. We also plan
to obtain higher resolution maps on this cloud and others with the
Arecibo Telescope to complement the lower resolution GBT data and
gain insight into the smaller scale structure. Further in the future,
we plan to apply the technique to deep, very high resolution observations
of such clouds using a large array such as the Australian Square Kilometre
Array Pathfinder (ASKAP).

\acknowledgements{}

It is our pleasure to thank Felix J. Lockman for his support in observations,
data reduction and interpretative discussions. We also thank Andrew
Fox, Joss Bland-Hawthorn, Bart Wakker, Jennifer Stone, Ellen Zweibel,
Alex Hill, Matt Haffner and Kelley Hess for stimulating discussions
as well as the anonymous referee for useful comments leading to improvements
to the paper. We acknowledge support by the NSF through grants AST-0908134
and AST-708967. LN thanks the Arecibo Observatory for a pre-doctoral
fellowship. SS acknowledges support by the Research Corporation for
Science Advancement. The Arecibo Observatory is operated by SRI International
under a cooperative agreement with the National Science Foundation
(AST-1100968), and in alliance with Ana G. M\'{e}ndez-Universidad Metropolitana,
and the Universities Space Research Association. The National Radio Astronomy Observatory is a facility of the National Science Foundation operated under cooperative agreement by Associated Universities, Inc.

\begin{flushleft}
 {\it Facilities:} \facility{GBT}.

\par\end{flushleft}

\begin{thebibliography}{}
\bibitem[Agertz et al.(2007)]{agertz07} Agertz, O., et al.\ 2007, \mnras, 380, 963
\bibitem[Begelman \& Fabian(1990)]{begelman90}Begelman, M.~C., \& Fabian, A.~C.~1990, \mnras, 244, 26P
\bibitem[Besla et al.(2007)]{besla07}Besla, G., Kallivayalil, N., Hernquist, L., Robertson, B., Cox, T.~J., van der Marel, R.~P., \& Alcock, C.~2007, \apj, 668, 949
\bibitem[Besla et al.(2010)]{besla10}Besla, G., Kallivayalil, N., Hernquist, L., van der Marel, R.~P., Cox, T.~J., \& Kere\v s, D.~2010, \apjl, 721, L97
\bibitem[Bland-Hawthorn et al.(2007)]{bh07} Bland-Hawthorn, J., Sutherland, R., Agertz, O., \& Moore, B.\ 2007, \apjl, 670, L109
\bibitem[Bland-Hawthorn(2009)]{bh09} Bland-Hawthorn, J.\ 2009, IAU Symposium, 254, 241
\bibitem[Brooks et al.(2009)]{brooks09} Brooks, A.~M.,  Governato, F., Quinn, T., Brook, C.~B.,  \& Wadsley, J.\ 2009, \apj, 694, 396  
\bibitem[Br\"{u}ns et al.(2001) ]{bruns01} Br\"{u}ns, C., Kerp, J., \& Pagels, A.\ 2001, \aap, 370, L26
\bibitem[Br\"{u}ns et al.(2005) ]{bruns05} Br\"{u}ns, C., et al.\ 2005, \aap, 432, 45
\bibitem[Collins et al.(2009)]{collins09} Collins, J.~A., Shull,  J.~M., \& Giroux, M.~L.\ 2009, \apj, 705, 962 
\bibitem[Connors et al.(2006)]{connors06}Connors, T.~W., Kawata, D., \& Gibson, B.~K.~2006, \mnras, 371, 108
\bibitem[Cowie \& McKee(1977)]{cowie77}Cowie, L.~L., \& McKee, C.~F.~1977, \apj, 211, 135 
\bibitem[Dalton \& Balbus(1993)]{dalton93}Dalton, W.~W., \& Balbus, S.~A.~1993, \apj, 404, 625
\bibitem[Dekel  \& Birnboim(2006)]{dekel05} Dekel, A., \& Birnboim, Y.\ 2006, \mnras, 368, 2  
\bibitem[Draine(2011)]{draine11}Draine, Bruce T.~2011, Physics of the Interstellar and Intergalactic Medium (Princeton, NJ, USA: Princeton University Press)
\bibitem[Elmegreen(1997)]{elmegreen97}Elmegreen, B.~G.~1997, \apj, 477, 196
\bibitem[Esquivel et al.(2006)]{esquivel06}Esquivel, A., Benjamin, R.~A., Lazarian, A., Cho, J., \& Leitner, S.~N.~2006, \apj, 648, 1043
\bibitem[Fang et al.(2006)]{fang06}Fang, T., Mckee, C.~F., Canizares, C.~R., \& Wolfire, M.~2006, \apj, 644, 174
\bibitem[Faucher-Gigu\`ere et al.(2009)]{faucher09}Faucher-Gigu\`ere, C.~A., Lidz, A., Zaldarriaga, M., \& Hernquist, L.~2009, \apj, 703, 1416 
\bibitem[Fisher et al.(2003)]{fisher02}Fisher, J.~R., Norrod, R.~D., \& Balser, D.~S.~2003 GBT Electronics Division Internal Report No.~312, http://www.gb.nrao.edu/electronics/edir/index.html
\bibitem[Fox et al.(2005)]{fox05}Fox, A.~J., Wakker, B.~P., Savage, B.~D., Tripp, T.~M., Sembach, K.~R., \& Bland-Hawthorn, J.~2005, \apj, 630, 332
\bibitem[Fox et al.(2010)]{fox10}Fox, A.~J., Wakker, B.~P., Smoker, J.~V., Richter, P., Savage, B.~D., \& Sembach, K.~R.~2010, \apj, 718, 1046
\bibitem[Gallagher  \& Smith(2005)]{gallagher05} Gallagher, J.~S., \& Smith, L.~J.\ 2005, Extra-Planar Gas, 331, 147 
\bibitem[Gnat et al.(2010)]{gnat10}Gnat, O., Sternberg, A., \& McKee, C.~F.~2010, \apj, 718, 1315
\bibitem[Heitsch \& Putman(2009)]{heitsch09}Heitsch, F., \& Putman, M.~E.~2009, \apj, 698, 1485
\bibitem[Hilditch et al.(2005)]{hilditch05}Hilditch, R.~W., Howarth, I.~D., \& Harries, T. J.~2005, \mnras, 357, 304
\bibitem[Jin \& Lynden-Bell(2008)]{jin08}Jin, S., \& Lynden-Bell, D.~2008, \mnras, 383, 1686
\bibitem[Kalberla  \& Haud(2006)]{kalberla06} Kalberla, P.~M.~W., \& Haud, U.\ 2006, \aap, 455, 481 
\bibitem[Kere{\v s} et al.(2005)]{keres05} Kere{\v s}, D.,  Katz, N., Weinberg, D.~H., \& Dav{\'e}, R.\ 2005, \mnras, 363, 2  
\bibitem[Koerwer(2009)]{koerwer09}Koerwer, J.~F.~2009, \aj, 138, 1
\bibitem[Kwak et al.(2011)]{kwak11}Kwak, K., Henley, D.~B., \& Shelton, R.~L.~2011, \apj, 739, 30 
\bibitem[Mastropietro et al.(2005)]{mastro05}Mastropietro, C., Moore, B., Mayer, L., Wadsley, J., \& Stadel, J.~2005, \mnras, 363, 509
\bibitem[Mathis et al.(2002)]{mathis02}Mathis, J.~S., Whitney, B.~A., \& Wood, K.~2002, \apj, 574, 812
\bibitem[McClure-Griffiths et al.(2009)]{mcclure09}McClure-Griffiths, N.~M., et al.~2009, \apjs, 181, 398
\bibitem[McKee \& Cowie(1977)]{mckee77} McKee, C.~F., \& Cowie, L.~L.\ 1977, \apj, 215, 213 
\bibitem[Moore \& Davis(1994)]{moore94}Moore, B., \& Davis, M.~1994, \mnras, 270, 209
\bibitem[Mori  \& Burkert(2001)]{mori01} Mori, M., \& Burkert, A.\ 2001, The Physics of Galaxy Formation, 222, 359 
\bibitem[Nidever et al.(2008)]{nidever08}Nidever, D.~L., Majewski, S.~R., \& Burton, W.~B.~ 2008, \apj, 679, 432
\bibitem[Nidever et al.(2010)]{nidever10}Nidever, D.~L., Majewski, S.~R., Butler Burton, W., \& Nigra, L.~2010, \apj, 723, 1618
\bibitem[Nelder \& Mead(1965)]{nelder65}Nelder and Mead, 1965, Computer Journal, Vol 7, pp 308-313.
\bibitem[Pittard(2007)]{pittard07}Pittard, J.~M.~2007, in Diffuse Matter from Star Forming Regions to Active Galaxies, ed. T.~W.~Hartquist et al.~(Dordrecht, The Netherlands: Springer)
\bibitem[Putman et al.(2003a)]{putman03a}Putman, M.~E., Staveley-Smith, L., Freeman, K.~C., Gibson, B.~K., \& Barnes, D.~G.\ 2003, \apj, 586, 170
\bibitem[Putman et al.(2003b)]{putman03b}Putman, M.~E., Bland-Hawthorn, J., Veilleux, S., Gibson, B.~K., Freeman, K.~C., \& Maloney, P.~R.~2003, \apj, 597, 948
\bibitem[Quilis \& Moore(2001)]{quilis01}Quilis, V., \& Moore, B.\ 2001, \apjl, 555, L95
\bibitem[Reynolds et al.(1998)]{reynolds98}Reynolds, R.~J., Tufte, S.~L., Haffner, L.~M., Jaehnig, K., \& Percival, J.~W.~1998, PASA, 15, 14
\bibitem[Sembach et al.(1995)]{sembach95}Sembach, K.~R., et al.\ 1995,
\bibitem[Sembach et al.(2003)]{sembach03}Sembach, K.~R., et al.\ 2003, \apjs, 146, 165
\bibitem[Shull et al.(2009)]{shull09} Shull, J.~M., Jones,  J.~R., Danforth, C.~W., \& Collins, J.~A.\ 2009, \apj, 699, 754 
\bibitem[Silk et al.(1987)]{silk87} Silk, J., Wyse, R.~F.~G.,  \& Shields, G.~A.\ 1987, \apjl, 322, L59 
\bibitem[Slavin et al.(1993)]{slavin93}Slavin, J.~D., Shull, J.~M., \& Begelman, M.~C.~1993, \apj, 407, 83
\bibitem[Spitzer(1998)]{spitzer98}Spitzer, Lyman, Jr.~1998, Physical Processes in the Interstellar Medium (Berlin, DE: Wiley-VCH)
\bibitem[Stanimirovi{\'c} et al.(2002)]{stani02}  Stanimirovi{\'c}, S., Dickey, J.~M., Kr{\v c}o, M.,  \& Brooks, A.~M.\ 2002, \apj, 576, 773 
\bibitem[Stanimirovi{\'c} et al.(2008)]{stani08}Stanimirovi\'{c}, S., Hoffman, S., Heiles, C., Douglas, K.~A., Putman, M., \& Peek, J.~E.~G.~ 2008, \apj, 680, 276
\bibitem[Stanimirovic et al.(2010)]{stani10} Stanimirovic, S.,  Gallagher, J.~S., III,  \& Nigra, L.\ 2010, Serbian Astronomical Journal, 180, 1 
\bibitem[T{\"u}llmann et  al.(2006)]{tullman06} T{\"u}llmann, R., Pietsch, W., Rossa, J., Breitschwerdt, D., \& Dettmar, R.-J.\ 2006, \aap, 448, 43 
\bibitem[Vieser \& Hensler(2007a)]{vieser07a}Vieser, W., \& Hensler, G.~2007a, \aap, 472, 141
\bibitem[Vieser \& Hensler(2007b)]{vieser07b}Vieser, W., \& Hensler, G.~2007b, \aap, 475, 251
\bibitem[Wakker et al.(2008)]{wakker08} Wakker, B.~P., York,  D.~G., Wilhelm, R., et al.\ 2008, \apj, 672, 298 
\bibitem[Weiner \& Williams(1996)]{weiner96}Weiner, B.~J., \& Williams, T.~B.~1996, \aj, 111, 1156
\bibitem[Wolfire et al.(1995)]{wolfire95b}Wolfire, M.~G., McKee, C.~F., Hollenbach, D., \& Tielens, A.~G.~G.~M.~1995, \apj, 453, 673 
\bibitem[Wood \& Loeb(2000)]{wood00}Wood, K. \& Loeb, A. 2000, \apj, 545, 86
\bibitem[Wood et al.(2005)]{wood05}Wood, K., Haffner, L.~M., Reynolds, R.~J., Mathis, J.~S., \& Madsen, G.~2005, \apj, 633, 295 
\bibitem[Zheng et al.(2002)]{zheng02}Zheng, Z., \& Miralda-Escud\'e, J.\ 2002, \apjl, 568, L71
\end{thebibliography}
\end{document}